\documentclass[11pt]{article}
\usepackage{epsfig}
\usepackage[usenames]{color}
\usepackage{amsmath}
\usepackage{amsfonts}
\usepackage{amssymb}
\usepackage{mathrsfs}
\numberwithin{equation}{section}

\newtheorem{proposition}{Proposition}[section]
\newcommand{\bpr}{\begin{proposition}}
\newcommand{\epr}{\end{proposition}}

\newcounter{Roman}
\addtocounter{Roman}{1}

\newcommand{\beq}{\begin{equation}}
\newcommand{\eeq}{\end{equation}}
\newcommand{\bea}{\begin{eqnarray}}
\newcommand{\eea}{\end{eqnarray}}
\newcommand{\bal}{\begin{align}}
\newcommand{\bml}{\begin{multline}}
\newcounter{saveeqn}

\newcommand{\ssc}{\scriptscriptstyle}

\newcommand{\tlepsilon}{\tilde{\epsilon}}
\newcommand{\wdelta}{\widehat{\delta}}
\newcommand{\wDelta}{\widehat{\Delta}}

\input epsf
\begin{document}  

\begin{center}{\Large\bf  Causality and Unitarity via the Tree-Loop Duality Relation}\\[2cm] 
{E. T. Tomboulis\footnote{\sf e-mail: tomboulis@physics.ucla.edu}
}\\[0.1cm]
{\em Mani L. Bhaumik Institute for Theoretical Physics\\
%and  \\
Department of Physics and Astronomy, UCLA, Los Angeles, 
CA 90095-1547} 
\end{center}
\vspace{1cm}

\begin{center}{\Large\bf Abstract}\end{center}
The tree-loop duality relation is used as a starting point to derive the constraints of causality and unitarity. Specifically, the Bogoliubov causality condition is ab initio derived at the individual graph level. It leads to a representation of a graph in terms of lower order cut graphs. Extracting the absorptive part gives then the general 
unitarity relation (Cutkosky rule). The derivation, being carried out  directly in momentum space, holds for  any local (polynomial) hermitian interaction vertices. This is in contrast to the technical difficulties arising from contact terms in the spacetime approach based on the largest time equation.

\vfill
\pagebreak

\section{Introduction} 
The tree-loop duality (TLD) relation \cite{Cetal}, \cite{BCDR} gives a representation of a generic loop graph in $d$ dimensions as a sum of trees integrated over  $(d-1)$-dimensional phase space, the trees arising from making all cuts with one cut per loop so as to open the diagram into a tree. 
The propagators in this representation have a modified (momentum dependent) $i\epsilon$ prescription. 
Reexpressing these propagators in terms of propagators with the standard $i\epsilon$ prescription leads back to the original Feynman tree-loop theorem \cite{F}, which includes all possible multiple cuts in each loop. 

Such relations between loops and trees are useful in a variety of contexts.  Since they are formulated directly in momentum space they offer an alternative avenue for the direct consideration of the singularity structure and the cancellation of UV and IR singularities in amplitudes, cf. \cite{BChDMR}, \cite{H-PSR}.  In \cite{ChBDR}, \cite{BChDR}   numerical implementation of LTD was introduced and applied in the evaluation of multi-leg amplitudes. 

In this paper we use the tree-loop relation to give ab initio derivations of the constraints of causality and unitarity at the individual graph level and in the presence of general local (polynomial) derivative interactions. 

 A general formulation of the causality requirement in quantum field theory is provided by the Bogoliubov causality condition (BCC) \cite{Bo}, \cite{BS}. This condition, formulated in terms of the S-matrix operator (evolution matrix), constraints amplitudes both in the timelike and spacelike region. It implies the usual `microcanonical' causality condition of commutativity of operators at spacelike separations as a special case. 
In the original treatment \cite{BS} the unitarity condition $SS^\dagger=1$ and the BCC equation 
(cf. (\ref{BCC0d}) below)  
are imposed as fundamental constraints on the $S$-matrix operator. Just these  two constraints are then shown to completely determine the  structure of the $S$-matrix operator in perturbation theory, i.e., that it is given by the time-ordered exponential of a suitable interaction Lagrangian.  

Most often, however, one is interested in the reverse problem, that is, given some Lagrangian and its associated Feynman rules to verify that the resulting amplitudes satisfy unitarity and causality.  
It turns out that the BCC by itself is rather powerful in this regard. If it holds for a given theory with a hermitian Lagrangian, then, the condition of unitarity turns out to be also satisfied. Furthermore, as noted above, the usual requirement of microcausality  is also satisfied as a special case of the general BCC equation. The BCC equation implies, in fact, an exact representation of a graph in terms of Cutkosky cuts separating the graph into lower order pieces. This representation, quite distinct from that of the Feynman tree-loop or TLD representations, has other applications as well, e.g., a method of implementing the iterative cancellation of UV subdivergences in perturbative renormalization \cite{TT1}. 

It should be noted here that, as it is well known, a general and rather elegant approach for deriving such results is provided by Veltman's largest time equation \cite{V}. 
This equation allows derivation of general cutting rules from which the BCC and the 
general unitarity relation can be obtained at the level of individual graphs and at all loop orders \cite{V} \cite{tHV}. This approach, however, runs into some technical difficulties in the presence of derivative interactions \cite{tHV}. Because it is initially formulated in coordinate space such interactions, acting on theta functions specifying time direction, 
generate undesired contact terms
%\footnote{In fact, the numerators of spin-1 and higher propagators also generate such terms and have to be dealt with %appropriately. \label{Fnprop}} 
spoiling straightforward derivation. Such contact terms could in principle be cancelled by appropriate non-covariant counterterms, but, for general interactions and multi-loop diagrams, this can quickly become unmanageable. The approach based on the TLD relation, on the other hand, though less elegant, is formulated directly in momentum space, where derivative interaction appear as polynomials (entire functions) in momenta, and does not encounter any such difficulties. 

We also note in this connection that, more recently,  another approach to causality formulated directly in terms of retarded propagators has been given in \cite{DFMC}. It may be obtained via cutting rules in the Schwinger-Keldysh in-in formalism \cite{KS}. The relation to the  BCC equation formulation is indicated in \cite{DFM}.

The paper is organized as follows. In section 2 we review TLD and its derivation. This serves mainly to establish our notations and present a variety of relations and identities involving different types of propagators that are needed in the sequel. In section 3 we briefly review the BBC and proceed to derive it from the TLD representation for 1-loop graphs. By combining both causal orderings of two particular vertices entering this equation we then obtain the BCC representation of a graph mentioned above. In section 4 we use this representation to obtain the unitarity relation by isolating the absorptive part. The microcausality condition is also shown to be implied by the BCC equations. In  section 5 we present the extension to multi-loops by way of a 2-loop case that illustrates some new features appearing in higher loops. Section 6 contains some concluding remarks.

\section{ The tree-loop relation\label{TL}} 
Consider a generic 1-loop graph with
$N$ vertices.  We denote by $p_i$ the total external momentum flowing into the $i$-th vertex, $i=1,\ldots, N$ (Fig. 1).  
\begin{figure}[ht]
\begin{center}
\includegraphics[width=\textwidth]{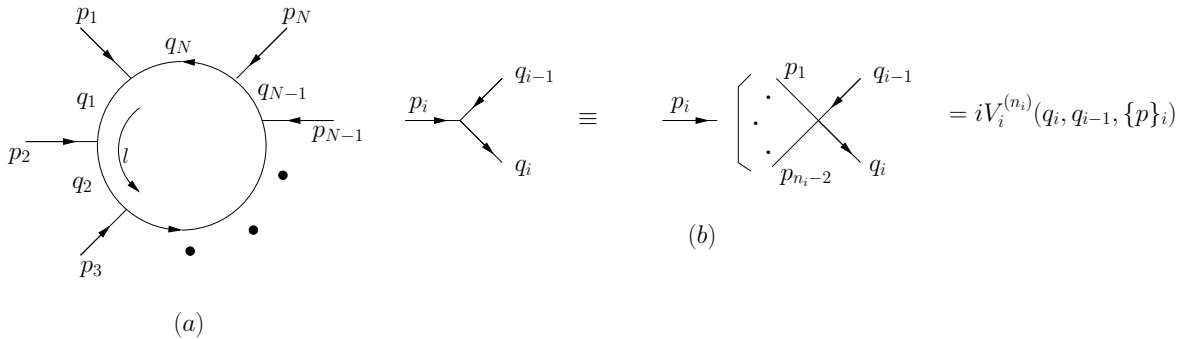}
\end{center}
\caption{(a) General 1-loop graph; (b) vertices depicted in abbreviated form as shown.      \label{tlF1}}
\end{figure}
All external momenta are taken to be incoming and anti-clockwise ordered as shown.  By momentum conservation we then have $\sum_{j=1}^{N} p_j = 0 $.
%\beq 
%\sum_{j=1}^{N} p_j = 0  %\, , \qquad \mbox{and} \qquad \sum_{i=1}^{n_V} \bp_i = 0 
%\, .    \label{momcon}
%\eeq
For a 1-loop graph the number of internal lines equals the number of vertices $N$. The internal line momenta, labeled $q_k = (q_{j\,0}, {\bf q}_k)$, then are  
\beq 
q_k = l +\sum _{j=1}^k p_j   \, , \qquad k=1,\ldots, N \, ,  \label{intmom1}
\eeq
where $l$ is the loop momentum taken to flow anti-clockwise. Note that%, by (\ref{momcon}), 
$q_N=l$.    
The amplitude for the general 1-loop graph in $d$  spacetime dimensions is then given by 
\beq 
A( \{p\}_{\ssc N}) = \int  \, {d^dl \over (2\pi)^d} \, \prod_{i=1}^{N} \, iV_i^{(n_i)}(q_i, q_{i-1}, \{p\}_i) 
 \, \Delta(q_i)  
  \, .   \label{ampl1}
\eeq
Here $ \{p\}_N$ labels the set of all external momenta. $\Delta(q)$ denotes the usual causal (Feynman) propagator:  
\beq
\Delta(k) =  {i  \over (k^2 - m^2 +i\epsilon) }  \,  . \label{prop1}
\eeq 
$V_i^{(n_i)}(q_i,q_{i-1}, \{p\}_i)$ denotes the $i$-th vertex factor 
with $n_i$ legs and with 
$\{p\}_i$ standing for the set of the $(n_i-2)$ external momenta, of total momentum $p_i$, flowing into the vertex (Fig. 1(b)). It is taken to be a real polynomial in its  momentum arguments. For notational simplicity we take scalar vertex factors, the (basically trivial) generalization to tensor $V$'s adding nothing new in the following. 
We assume that (\ref{ampl1}) is appropriately regulated so that it is UV  finite. Dimensional regularization or, alternatively,  Pauli-Villars (PV) regularization  may be used; in the latter case the appropriate number of subtractions in (\ref{ampl1}) with the mass $m$ replaced by the regulator masses is implicitly assumed.

We proceed to carry out the integration over the time component $l_0$ of the loop momentum by closing the integration contour at infinity. This, of course, assumes that the integrand is sufficiently convergent to give vanishing contribution from the contour at infinity. This, in general, is the case if the regulated integral (\ref{ampl1}) is UV finite. The integrand in (\ref{ampl1}) possesses poles at 
\beq 
q_{k\, 0}^{\pm} = \pm \sqrt{\omega_{{\bf q}_k}^2 -i\epsilon} \qquad \mbox{ with} \qquad  \omega_{{\bf q}_k} = +\sqrt{{\bf q}_k^2 + m^2}  \; . \label{poles1}
\eeq 
We close the contour in the $l_0$ lower half-plane (l.h.p.). 
At the l.h.p.  pole $q_{k\,0}^+$ of the k-th internal line one then finds 
\beq
 \left[{\rm Res}\left(  \Delta(q_k) \right)\right]_{\rm k-th\  pole}   =    \, \frac{i}{2  \omega_{{\bf q}_k} }  
   =  i \int d l_0 \,  \theta(q_{k\, 0}) \delta(q_k^2 - m^2)   \label{Res2}
 \eeq
and 
\beq 
\left[ q_j^2 - m^2 +i\epsilon\right]_{q^+_{k\, 0}} = [q_j^2 - m^2 -i\epsilon (q_{j\, 0} - q_{k\, 0})/ \omega_{{\bf q}_k} ]_{q_{k\,0}=\omega_{{\bf q}_k}}
  \, .   \label{Res3}
\eeq 
Note that $(q_j - q_k)$ is a linear combination of only the external momenta. Similarly, for the arguments of $V_i^{(n_i)}$ at the k-th pole one sets 
\beq 
q_j|_{q_{k\,0}=\omega_{{\bf q}_k}} = [q_k + (q_j - q_k)] _{q_{k\,0}=\omega_{{\bf q}_k}}  \label{Res4}
\eeq
since, the vertex functions being entire, the $\epsilon$'s are irrelevant. 
Combining (\ref{Res2}), (\ref{Res3}) and (\ref{Res4}) the residue theorem applied to (\ref{ampl1}) gives 
\beq 
 A(\{p\}_{\ssc N})  =   \sum_{k=1}^{N}   \int  \, {d^dl \over (2\pi)^d} \;\;
\wdelta^+(q_k)      
 \,  \prod_{i=1}^{N} \, i V_i^{(n_i)}(q_i, q_{i-1}, \{p\}_i) \,
  \prod_{j=1 \atop j \not=k}^{N} \,   \Delta\big(q_j, \tlepsilon(q_j,q_k)\big) \, , \qquad\qquad    \label{tld} 
\eeq
where we introduced the notations 
  \beq  
\Delta\big(q_j, \tlepsilon(q_j,q_k)\big) \equiv {i\over q_j^2 - m^2 - i\epsilon (q_{j\,0}-q_{k\, 0})/\omega_{{\bf q}_k} }  \, \label{prop2}
  \eeq
  and 
  \beq 
 \wdelta^{\pm}(k) \equiv 2\pi \delta^{\pm}(k^2-m^2) = 2\pi \theta(\pm k_0) \delta(k^2-m^2)  \,.  \label{deltaplusmin} 
  \eeq
We note in passing that in (\ref{prop2}) $\omega_{{\bf q}_k}$, being positive, may be absorbed into the infinitesimal $\epsilon$, i.e., one may replace $\epsilon/\omega_{{\bf q}_k} \to \epsilon$. The factor  $(q_{j\,0}-q_{k\, 0})$, on the other hand, which can be of either sign, is crucial for locating the poles in (\ref{prop2}). One may also write $i\epsilon (q_{j\,0}-q_{k\, 0})$ in a covariant form by introducing an auxiliary time-like vector \cite{Cetal} but this will not be needed for our purposes here. 

(\ref{tld}) represents the original 1-loop amplitude as a sum over single cuts that convert it to a sum of trees integrated over one-body phase space. It has therefore been referred to in the literature by the somewhat ponderous name of ``loop-tree duality theorem". 
The result at 1-loop level is actually elementary, the only subtlety involved being keeping track of the $i\epsilon$'s in evaluating a propagator at the pole of another propagator.  Nonetheless, this loop-to-tree reduction, and its higher loop extension, turns out to be rather useful in several contexts as noted above. 
It should be pointed out in this connection that the cut introduced %by $\wdelta^{+}(k)$ 
in (\ref{tld}) is indeed a standard Cutkosky cut (Fig. 2). 
\begin{figure}[ht]
\begin{center}
\includegraphics[width=0.6\textwidth]{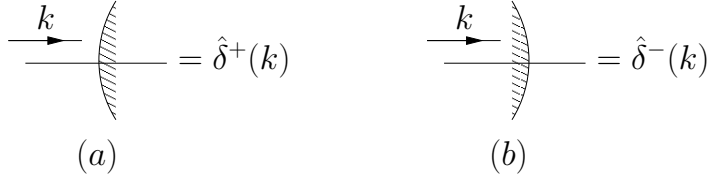}
\end{center}
\caption{Graphical depiction of Cutkosky cuts: (a) positive; (b) negative. Energy flows form the unshaded to the shaded side.   \label{tlF2}}
\end{figure}

The propagator  (\ref{prop2}) is related to the one with the usual $i\epsilon$ prescription by the 
relation\footnote{(\ref{proprel1}) - (\ref{proprel1c}) and (\ref{ARprop2}) below are straightforwardly  obtained  from the  Plemelj-Dirac formula: 
\[ {1\over x \mp i\epsilon} = {\rm P} \left( {1\over x}\right)  \pm i\pi \delta(x)   \, . \] \label{F-PD}}
\beq 
  \Delta\big(q_j, \tlepsilon(q_j,q_k)\big) = \Delta(q_j)  - 2\pi \theta(q_{j\,0} - q_{k\,0}) \delta(q_j^2 - m^2) \, . \label{proprel1}
  \eeq
Similarly, on has  
\beq 
  \Delta\big(q_j, \tlepsilon(q_j,q_k)\big) =   -\Delta^*(q_j)  + 2\pi \theta(-(q_{j\,0} - q_{k\,0})) \delta(q_j^2 - m^2)   \, .  \label{proprel1a} 
  \eeq 
Since $\tlepsilon(q_j,q_k)= - \tlepsilon(q_k,q_j)$ one can also write  
\bea
 \Delta\big(q_j, -\tlepsilon(q_j,q_k)\big) & = & \Delta(q_j)  - 2\pi \theta(-(q_{j\,0} - q_{k\,0})) \delta(q_j^2 - m^2) \label{proprel1b}   \\
  \Delta\big(q_j, -\tlepsilon(q_j,q_k)\big) & = & -\Delta^*(q_j)  + 2\pi \theta(q_{j\,0} - q_{k\,0}) \delta(q_j^2 - m^2) \,.\label{proprel1c}
\eea
The relations (\ref{proprel1}) - (\ref{proprel1c}) will be useful for us in the sequel. 

Inserting (\ref{proprel1}) in (\ref{tld}) and expanding one regains the original Feynman tree-loop  relation which expresses a loop integral as a sum over up to $N$ cuts. (In physical amplitudes some of these multiple cuts may give zero contribution for kinematical reasons.) The relation to the Feynman tree theorem is extensively discussed in \cite{Cetal}.  

It is also very useful to consider the advanced and retarded propagators given, respectively, by:  
\beq
\Delta_A(q) = {i\over q^2 - m^2 -i\epsilon q_0}  \;, \qquad  \Delta_R(q) = {i\over q^2 - m^2 +i\epsilon q_0}   \, . \label{ARprop1}  \\
\eeq
$\Delta_R(q)$ has poles only in the l.h.p., whereas $\Delta_A(q)$ has poles only in the upper half plane (u.h.p.).  One has the relations 
\beq
\Delta_A(q) = - \Delta_R^*(q) = \Delta_R(-q)       \label{ARprop2}
\eeq
and 
\beq
\Delta_R(q) = \Delta(q) - \wdelta^-(q)\, , \qquad \Delta_A(q) = \Delta(q) - \wdelta^+(q)   \, .      \label{ARprop3}
\eeq 
In fact, the original Feynman derivation is based on the relation (\ref{ARprop3}). 

Let $Q$ denote a set of internal lines. 
In the following we will conveniently use the label $Q$ for both a set of lines and the set of internal momenta $\{ q_1, q_2,\ldots\}$ they carry. 
We also introduce the notations 
\beq
\Delta(Q) \equiv \prod_{j\in Q} \Delta(q_j) \, , \quad\quad   \Delta_{A(R)}(Q) \equiv \prod_{j\in Q} \Delta_{A(R)}(q_j)   
%\quad\quad     \wDelta  \equiv \sum_{k\in Q} \wdelta^+(q_k) \,\prod_{j\in Q\atop j\not= k} \Delta(q_j, \tlepsilon(q_j,q_k)) 
\label{prodnot1}
\eeq
and 
\beq
\wDelta^{\pm}(Q)  \equiv \sum_{k\in Q} \wdelta^{\pm}(q_k) \,\prod_{j\in Q\atop j\not= k} \Delta(q_j, \pm\tlepsilon(q_j,q_k)) \, . \label{prodnot2}
\eeq
Then 
\beq
\Delta_A(Q) = \Delta(Q) - \wDelta^+(Q)  \, . \label{Gproprel1}
\eeq
This rather non-trivial relation was proved in \cite{BCDR}. 
From (\ref{ARprop2}) one then also has 
\beq
\Delta_R(Q) = \Delta(Q) - \wDelta^-(Q)  \, . \label{Gproprel2}
\eeq
(\ref{Gproprel1}) and (\ref{Gproprel2}) reduce, of course, to (\ref{ARprop3}) in the special case of 
$Q$ consisting of one line. 

Further very useful relations follow \cite{BCDR} by iterating (\ref{Gproprel1}) and (\ref{Gproprel2}). Thus, consider  
\bea
\wDelta^+(Q_1\cup Q_2) & = & -\Delta_A(Q_1\cup Q_2) + \Delta(Q_1\cup Q_2) = -\prod_{i=1,2} \Delta_A(Q_i) +
\prod_{i=1,2} \Delta(Q_i)  \nonumber \\ 
& = & -\prod_{i=1,2} [\Delta(Q_i) - \wDelta^+(Q_i)] + \prod_{i=1,2} \Delta(Q_i) 
\eea
where (\ref{Gproprel1}) was used in the first line and then once again in the second line. Expanding one obtains   
\beq
\wDelta^+(Q_1\cup Q_2)  = - \wDelta^+(Q_1)\wDelta^+(Q_2) + \wDelta^+(Q_1) \Delta(Q_2)  + \Delta(Q_1) \wDelta^+(Q_2) \, . \label{Gproprel3}
\eeq
This may be obviously extended to the union of any family $Q_1, \cdots, Q_n$ of sets of lines. 

Employing (\ref{Gproprel1}) allows a short derivation of the loop-tree relation (\ref{tld}) close in spirit to the original Feynman argument \cite{F}.  Replace in (\ref{ampl1}) each factor $\Delta(q_i)$ by 
$\Delta_A(q_i)$, and proceed as before to carry out the integration over $l_0$  by closing the contour in the l.h.p. Since $\Delta_A(q)$ has no poles in the l.h.p. the result of the integration is now equal to zero: 
\beq 
\int  \, {d^dl \over (2\pi)^d} \, \prod_{i=1}^{N} \, iV_i^{(n_i)}(q_i, q_{i-1}, \{p\}_i) 
 \, \prod_{j=1}^{N} \Delta_A(q_j)    = 0   \, . \label{Altd}
 \eeq
Using (\ref{Gproprel1}) in (\ref{Altd}), with the set $Q$ consisting of all $N$ internal lines, one immediately obtains (\ref{tld}). 

Extension of the tree-loop duality relation to two and higher loop order \cite{BCDR} gives a representation of a graph in terms of cuts that open the multiloop diagram into a tree integrated over multiple-body phase space. (\ref{Gproprel3}) and its generalizations are useful in this extension (cf. section \ref{ml}). 
In our application to BCC and unitarity below 1-loop diagrams already result into a 2-loop tree-loop relation. 

%finiteness - divergences remarks

\section{Causality  \label{caus}}
\subsection{The Bogoliubov causality condition \label{bccsubs}}  
A general statement of the requirement of causality in quantum field theory is provided by the BCC \cite{Bo} \cite{BS}. 
The BCC is arrived at by considering a spacetime interaction region $G$ split into two complimentary subregions $G_1$ and $G_2$ separated by a space-like surface at some fixed time $t$; points in $G_2$ are at later and points in $G_1$ at earlier times than $t$. The strength of the interaction may be varied by replacing the interaction Lagrangian $ {\cal L}$ by $g(x){\cal L}$, where $g(x)$ is a real function in the range $[0,1]$. The notion of causality in then introduced by requiring that variations of $g(x)$ in $G_2$ do not affect the dynamics in $G_1$. This requirement is expressed in terms of the $S$-matrix operator (evolution operator) and amounts to its factorization between $G_2$ and $G_1$. Now, for two spacetime events $x,y$, the statement $y^0 > x^0$ translates into two relativistically invariant possibilities: either $y$ is in the causal future of $x$ (in or on the future light cone of $x$): $x \prec y$; or $x$ and $y$ are separated by a spacelike interval: $x\sim y$. With $S=S[g]$ as a functional of $g(x)$ then, the above causality requirement can be stated in the form \cite{BS} 
\beq 
{\delta \over \delta g(x)} \left( {\delta  S \over \delta g(y)}  S^\dagger \right) =0 \, , \qquad x \precsim y  \, .\label{BCC0a}
\eeq
This may be written in the equivalent form 
\beq 
{\delta \over \delta g(y)} \left( S^\dagger {\delta  S \over \delta g(x)}  \right) =0 \, , \qquad x \precsim y  \, ,\label{BCC0b}
\eeq
by use of the relation 
%The $S$matrix is assumed to be unitary: $SS^\dagger =1$, which implies 
\beq 
{\delta  S \over \delta g(x)}  S^\dagger = - S{\delta  S^\dagger  \over \delta g(x)}   \, \label{BCC0c} 
\eeq
implied by the unitarity condition $SS^\dagger = 1$.
(\ref{BCC0b}) is more convenient for us in the following. Explicitly then, 
\beq 
S^\dagger {\delta^2  S \over \delta g(x) \delta g(y)}  + {\delta  S^\dagger \over \delta g(y)}{\delta  S \over \delta g(x)} = 
0 \, , \qquad x \precsim y  \, .\label{BCC0d}
\eeq
It is also useful to note here the hermitian conjugate of (\ref{BCC0d}) which expresses the causality condition 
interchanging $S$ with $S^\dagger$ amplitudes: 
\beq 
 {\delta^2  S^\dagger \over \delta g(x) \delta g(y)} S + {\delta  S^\dagger \over \delta g(x)}{\delta  S \over \delta g(y)} = 
0 \, , \qquad x \precsim y  \, .\label{BCC0e}
\eeq
The differentiations w.r.t. $g$ pick out two interaction vertices to be time-ordered, after which one may set $g(x)=1$.  
Matrix elements of (\ref{BCC0d}) between given initial and final states may then be computed to any order in perturbation theory  after imposing the time ordering by a theta-function insertion, multiplying with the appropriate external leg wave functions and passing over to momentum space. 

It should be noted that (\ref{BCC0d}) is more general than the  usual ``microcausality" requirement of commutativity of operators at spacelike separations. In contrast to the latter BCC constraints dynamics both in the timelike and the spacelike region, and, indeed, implies microcausality. %as a particular application of (\ref{BCC0d}). 
To extract the microcausality condition 
subtract (\ref{BCC0e}) from (\ref{BCC0d}), then, with reversed time ordering, subtract (\ref{BCC0d}) and add (\ref{BCC0e}).  
%add (\ref{BCC0d}) and (\ref{BCC0e}) but with reversed time ordering in the latter, i.e., $x$ and $y$ interchanged. From this sum subtract the sum of (\ref{BCC0d}) with time ordering reversed and  (\ref{BCC0e}). 
One then obtains 
\begin{align}
& \left(S^\dagger {\delta^2  S \over \delta g(x) \delta g(y)}  - {\delta^2  S^\dagger \over \delta g(x) \delta g(y)} S\right) \theta(y^0-x^0) - 
 \left(S^\dagger {\delta^2  S \over \delta g(y) \delta g(x)}  - {\delta^2  S^\dagger \over \delta g(y) \delta g(x)} S\right) \theta(x^0-y^0) \nonumber  \\
 =& - \left( {\delta S^\dagger \over \delta g(y)} {\delta S\over \delta g(x)} - 
{\delta S^\dagger\over \delta g(x)} {\delta S \over \delta g(y)}  \right)     \label{BCC0mc1} 
\end{align} 
%Now from the usual considerations of relativistic invariance one straightforwardly deduces \cite{BS}  that, 

Now, for $x\sim y$ the l.h.s. of (\ref{BCC0mc1}) vanishes it is easily seen\footnote{Under a Lorentz transformation $x\to Lx$, $g$ transforms as 
$g^\prime(x) = g(L^{-1}x)$ (i.e., a scalar function), and $S[g^\prime] = U_L S[g]U_L^\dagger$ 
where $U_L$ is the corresponding unitary operator transforming the field states \cite{BS}. \label{FLT}}  by going from a frame where $(y^0-x^0) > 0$ to a frame where $(y^0-x^0) < 0$. Thus,  
\beq
\left( {\delta S^\dagger \over \delta g(y)} {\delta S\over \delta g(x)} - 
{\delta S^\dagger\over \delta g(x)} {\delta S \over \delta g(y)}  \right)  = 0  \, , \qquad x\sim y \; .    \label{BCC0mc2} 
\eeq 
This is the statement of microcausality. 
%the commutativity of insertions of the interaction Lagrangian operator separated by spacelike distances. 

As mentioned in section 1 above, in the original treatment \cite{BS} the unitarity condition $SS^\dagger=1$ and the causality condition (\ref{BCC0d}) are imposed as fundamental constraints on the $S$-matrix operator, which then determine its structure in perturbation theory. 
In practice, however, one is usually interested in verifying that Feynman rules for some particular theory satisfy these constraints. The formulation via the BCC is particularly well-suited for this. This is due to the flexibility afforded through the introduction of the variable coupling device. This allows picking out all or some subset of the interaction terms in the Lagrangian or even individual insertions of certain vertices in the perturbative expansion. This implies that the conditions can be stated even in terms of individual graphs, as realized in \cite{V}.

\subsection{Derivation of BCC via tree-loop duality } 
Our goal here then is  to derive (\ref{BCC0d}) ab initio using tree-loop duality. 
Consider a 1-loop contribution to a S-matrix amplitude given by the  general 1-loop graph  (\ref{ampl1}), Fig. \ref{tlF1}. All external moment $p_i$ are now on-shell with a subset  $\{ p_1, \ldots, p_r\}$ taken as incoming and a subset $\{ p_{r+1}, \ldots, p_N\}$ as outgoing: $\sum_i^r p_i - \sum_{r+1}^N  p_i  = 0  $. 
%\beq \sum_i^r p_i - \sum_{r+1}^N  p_i  = 0  \,  . \label{momcon1}
%\eeq
Correspondingly, the internal momenta $q_i$ are split into two sets: $Q_I = \{ q_1, \ldots, q_r\}$ and $Q_{II}= \{ q_{r+1}, \ldots q_N\}$. We use the labels $Q_I, Q_{II}$ to denote the two sets of internal momenta as well as the corresponding internal lines. 

\begin{figure}[htb]
\begin{center}
\includegraphics[width=0.6\textwidth]{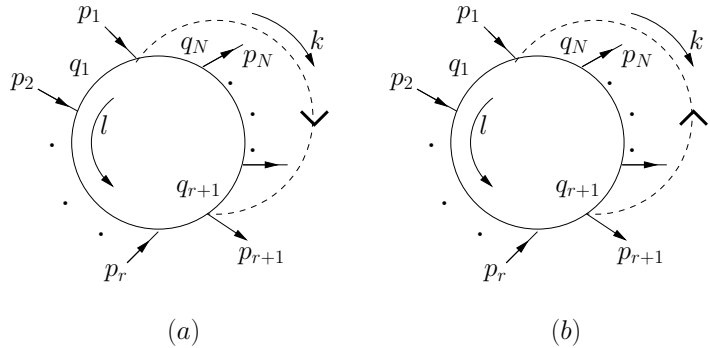}
\end{center}
\caption{(a) S-matrix amplitude with ordered pair of vertices entering the BCC equation (see text); (b) the same amplitude with opposite ordering. \label{tlF3}}
\end{figure}

\begin{figure}[ht]
\begin{center}
\vspace{0.3cm}
\includegraphics[width=0.8\textwidth]{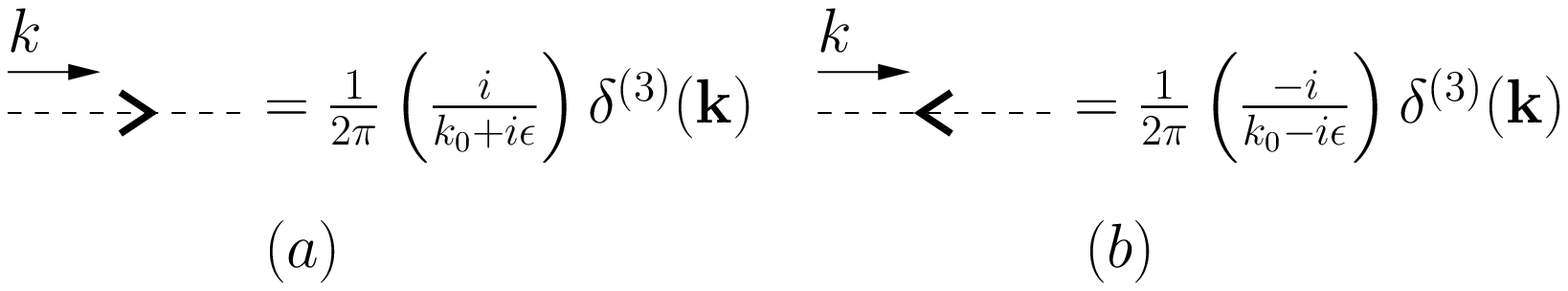}
\end{center}
\caption{Propagator connecting ordered pair of vertices; (a) and (b) depicting the two possible orderings.   \label{tlF4}}
\end{figure}

To derive the BBC equation for such an amplitude we consider the related amplitude (Fig. \ref{tlF3} (a)): 
\beq 
I_+( \{p\}_{\ssc N}) =   \int {d^dl \over (2\pi)^d}{d^dk \over (2\pi)}\,  \delta^{(d-1)}({\bf k}) \; {i\over k_0 + i\epsilon} \, \prod_{i=1}^{N} \, iV_i^{(n_i)}(q_i, q_{i-1}, \{p\}_i) 
 \, \Delta(q_i)  
  \, .   \label{Campl1}
\eeq
Here  the internal momenta are 
\beq 
q_i = l +\sum _{j=1}^i p_j \equiv l+ P_i \, ,\quad i\in Q_I\, ;   \qquad 
q_i = l + k + \left(\sum _{j=1}^r p_j  -   \sum _{j=r+1}^i p_j\right)  \equiv l+ k +  P_i \, ,\quad i\in Q_{II}\,  ,  \label{intmom2}
\eeq
with the loop momentum $l$ taken to flow anti-clockwise. Note that %, by (\ref{momcon1}), 
$q_N=l+k$. 
The additional (non-covariant) propagator in (\ref{Campl1})
%, serving to reproduce the time ordering in (\ref{BCC0d}), 
is defined in Fig. \ref{tlF4}. 
The $1/2\pi$ normalization of the $d^dk$ integration is left after cancellation 
of the usual $(1/2\pi)^d$  against the corresponding $(2\pi)^{d-1}\delta^{(d-1)}({\bf k})$ normalization. 

We now proceed to apply the tree-loop result (\ref{tld})  to the amplitude (\ref{Campl1}), which means an application to %essentially 
a 2-loop case. 
We first carry out the $k_0$ integration in (\ref{Campl1}) by closing the $k_0$-contour in the u.h.p. Only the poles in $\Delta(q_i)$ with $i\in Q_{II}$ then contribute. One may straightforwardly  evaluate their contribution in the analogous manner to (\ref{Res2}) - (\ref{tld}). Equivalently, and more concisely, one may proceed in analogy to (\ref{Altd}). In the present case one makes use of (\ref{Gproprel2}) by replacing all $\Delta(q_i)$ in (\ref{Campl1}) by $\Delta_R(q_i)$ if $i\in Q_{II}$.  Closing the contour in the u.h.p. now gives zero since no poles are enclosed. Substituting (\ref{Gproprel2}) then gives 
\bml
I_+( \{p\}_{\ssc N}) =  \sum_{n\in Q_{II}}   \int  {d^dl \over (2\pi)^d}{d^dk \over (2\pi)}\, \delta^{(d-1)}\; ({\bf k}) \;   \wdelta^-(q_n)\, {i\over k_0 + i\epsilon}  \\
\cdot  \prod_{s=1}^{N} iV_s^{(n_s)}(q_s, q_{s-1}, \{p\}_i) 
 \, \prod_{j\in Q_{II}\atop j\not= n}  \Delta(q_j, -\tlepsilon(q_j,q_n))  
  \prod_{i\in Q_{I}}  \Delta(q_i )  
  \, .   \label{Campl2}
\end{multline} 
In each term in the sum in (\ref{Campl2}) we have a $\wdelta^-(q_n)$ cut (cf. Fig. \ref{tlF2}) on a line $n \in Q_{II}$. The internal lines $j\in Q_{II}$, $j\not= n$ are split into two sets, those on the shaded side of the cut with $j< n$, and those on the unshaded side with $j>n$. 
 
Consider first the shaded side, $j< n$. 
%On the shaded side, %of the cut on line $n$ 
%Since $j < n$, one has, 
From  (\ref{intmom2}) one has $q_{j0} - q_{n0} = \sum_{i=j}^n p_{i0} > 0$. 
%\begin{itemize}\item 
Hence, starting with the first line, $j=n-1$, replacing the 
propagator $\Delta\big(q_j, -\tlepsilon(q_j,q_n)\big)$ in (\ref{Campl2}) by the relation (\ref{proprel1c}) we get two terms: one with propagator $(-\Delta^*(q_{n-1}))$; and another with a cut  $2\pi \delta(q_{n-1}^2 - m^2)= 
\wdelta^+(q_{n-1}) + \wdelta^-(q_{n-1})$. 
Now, in the latter term, it is easily seen that, when combined with the $\wdelta^-(q_n)$ cut in (\ref{Campl2}), the $\wdelta^+(q_{n-1}) $ contribution is kinematically allowed but, for stable particles, the $\wdelta^-(q_{n-1})$ contribution is not (cut lines and external legs (here outgoing) are on shell). Thus, in the presence of the $\wdelta^-(q_n)$ cut in (\ref{Campl2}) we have: 
%  \Delta\big(q_j, -\tlepsilon(q_j,q_k)\big) & = & -\Delta^*(q_j)  + 2\pi \theta(q_{j\,0} - q_{k\,0}) \delta(q_j^2 - m^2) \,.\label{proprel1c}
\bml
  \wdelta^-(q_n) \!\! \prod_{j\in Q_{II}\atop j < n}  \Delta(q_j, -\tlepsilon(q_j,q_n))  = \wdelta^-(q_n) \,   (-\Delta^*(q_{n-1})) \prod_{j\in Q_{II}\atop j < (n-1)}  \Delta(q_j, -\tlepsilon(q_j,q_n))        \\
 +    \wdelta^-(q_n) \, \wdelta^+(q_{n-1})  \!\! \prod_{j\in Q_{II}\atop j <(n-1)}  \Delta(q_j, -\tlepsilon(q_j,q_n))  \, . 
 \label{Cprod1} 
\end{multline}
%\end{itemize}
In the second term for each factor in the product, i.e., for each propagator $ \Delta(q_j, -\tlepsilon(q_j,q_n))$ on the unshaded side of the $\wdelta^+(q_{n-1})$ cut, we now use the relation (\ref{proprel1b}). Since, as noted above, $q_{j0} - q_{n0} >0$ for $j<n$ the second term becomes 
\beq 
\wdelta^-(q_n) \, \wdelta^+(q_{n-1}) \!\! \prod_{j\in Q_{II}\atop j <(n-1)}  \Delta(q_j, -\tlepsilon(q_j,q_n)) = 
\wdelta^-(q_n)  \wdelta^+(q_{n-1}) \!\! \prod_{j\in Q_{II}\atop j <(n-1)}  \Delta(q_j)
 \, .   \label{Cprod2}    
 \eeq  
We can now apply this reduction process to the remaining product of $\Delta(q_j, -\tlepsilon(q_j,q_n))$ propagators in the first term in (\ref{Cprod1}). Iterating we can thus reduce the l.h.s. of (\ref{Cprod1}) to only products of $\Delta(q_j)$ and $\Delta^*(q_j)$'s: 
\bml
 \wdelta^-(q_n) \!\!\prod_{j\in Q_{II}\atop j < n}  \Delta(q_j, -\tlepsilon(q_j,q_n))  = \wdelta^-(q_n) \!\! \prod_{j\in Q_{II}\atop j < n}(-\Delta^*(q_j) )    \\    
 +     \sum_{m\in Q_{II} \atop m<n}   \wdelta^-(q_n) \, \wdelta^+(q_m) \!\! \prod_{j\in Q_{II}\atop n> j > m}  (-\Delta^*(q_j) )  
 \prod_{k\in Q_{II}\atop m>k }  \Delta(q_k)  \, .   \label{Cprod3}    
 \end{multline}  
 
Going back to (\ref{Campl2})   consider next the unshaded side of the $\wdelta^-(q_n)$ cut. We now have $j>n$, and thus 
$q_{j0} - q_{n0} = -\sum_{i=n}^j p_{i0} < 0$. Using the relation (\ref{proprel1b}), we again get  
$2\pi \delta(q_j^2 - m^2)= \wdelta^+(q_j) + \wdelta^-(q_j)$ terms. In this case, however, when combined with the 
$\wdelta^-(q_n)$ factor already present in (\ref{Campl2}), both $\wdelta^{\pm}(q_j)$ contributions are easily seen to be not kinematically allowed by energy and momentum conservation constraints. 
Thus, we can write 
\beq
\wdelta^-(q_n) \!\prod_{j\in Q_{II}\atop j > n}  \Delta(q_j, -\tlepsilon(q_j,q_n)) = \wdelta^-(q_n)\!\prod_{j\in Q_{II}\atop j > n}  \Delta(q_j) \, . \label{Cprod4}
\eeq

Combining (\ref{Cprod3}) and (\ref{Cprod4}) and substituting in (\ref{Campl2}) we obtain
%\bml
%\wdelta^-(q_n)  \prod_{j\in Q_{II}\atop j\not= n}  \Delta(q_j, -\tlepsilon(q_j,q_n))  = 
%\wdelta^-(q_n) 
%\prod_{j\in Q_{II}\atop j > n}  \Delta(q_j)  \prod_{j\in Q_{II}\atop j < n}(-\Delta^*(q_j) )   \\
%   +     \sum_{m\in Q_{II} \atop m<n}  \wdelta^-(q_n) \,  \wdelta^+(q_m) \!\! \prod_{j\in Q_{II}\atop j > n}  \Delta(q_j) \prod_{j\in Q_{II}\atop n> j > m}  (-\Delta^*(q_j)  \;   
% \prod_{k\in Q_{II}\atop m>k }  \Delta(q_k)  \, . \label{Cprod4}
% \end{multline}
%Substituting (\ref{Cprod4}) into (\ref{Campl2}) we now obtain 
\begin{subequations} \label{Campl3}
\begin{align}
I_+( \{p\}_{\ssc N}) 
%& = & I_{\rm 1-cut}( \{p\}_{\ssc N}) + I_{\rm 2-cut} ( \{p\}_{\ssc N})  \nonumber \\
&=  \sum_{n\in Q_{II}} I^{(n)}_{\ssc +}( \{p\}_{\ssc N})   + \sum_{n,m \in Q_{II}\atop n>m} 
I^{(n,m)}_{\ssc +}( \{p\}_{\ssc N})   \,  \label{Campl3a} \\
&= \parbox{12cm}
{\epsfysize=2.2cm \epsfxsize=9.5cm \epsfbox{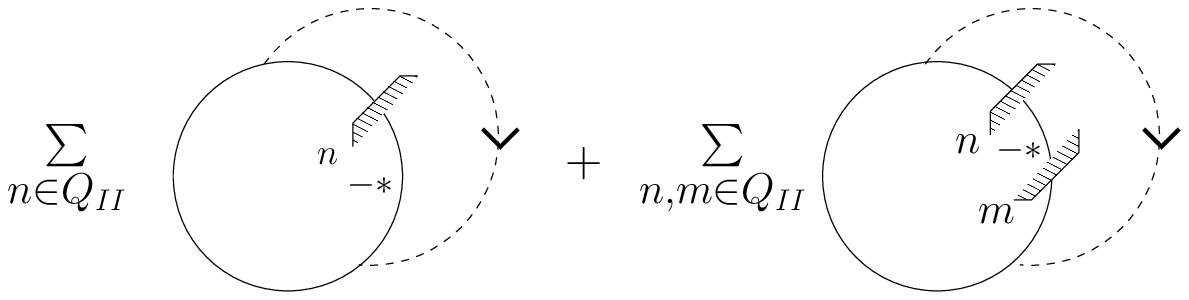}}     \label{Campl3b}
\end{align}
\end{subequations}
where 
\bml
I^{(n)}_{\ssc +}( \{p\}_{\ssc N}) = 
  \int  {d^dl \over (2\pi)^d}{d^dk \over (2\pi)}\, \delta^{(d-1)}\; ({\bf k}) \;  \wdelta^-(q_n) \;  {i\over k_0+ i\epsilon}   \\
\cdot  \prod_{s=1}^{N} iV_s^{(n_s)}(q_s, q_{s-1}, \{p\}_i)  
 \, \prod_{j\in Q_{II}\atop j > n}  \Delta(q_j)  \prod_{j\in Q_{II}\atop j < n}(-\Delta^*(q_j) )   
 \prod_{i\in Q_{I}}  \Delta(q_i )    \label{C1-cut1}
\end{multline} 
and 
\bml
I^{(n,m)}_{\ssc +}( \{p\}_{\ssc N}) = 
  \int  {d^dl \over (2\pi)^d}{d^dk \over (2\pi)}\, \delta^{(d-1)}\; ({\bf k}) \;  \wdelta^-(q_n) \;\wdelta^+(q_m) \;  {i\over k_0+ i\epsilon}   \\
\cdot  \prod_{s=1}^{N} iV_s^{(n_s)}(q_s, q_{s-1}, \{p\}_i) 
 \!\! \prod_{j\in Q_{II}\atop j > n}  \Delta(q_j) \prod_{j\in Q_{II}\atop n> j > m}  (-\Delta^*(q_j) ) \;   
 \prod_{k\in Q_{II}\atop m>k }  \Delta(q_k)   \prod_{i\in Q_{I}}  \Delta(q_i )   \, . \label{C2-cut1}
 \end{multline}
(\ref{Campl3b}) gives a graphical representation of $I^{(n)}_{\ssc +}( \{p\}_{\ssc N})$ and $I^{(n,m)}_{\ssc + }( \{p\}_{\ssc N})$. To avoid clutter all external lines are not explicitly shown. The label ``$\, -^*\,$" indicates that propagators on the shaded side of cuts in $Q_{II}$ are given by $-\Delta^*$. 

Note that in the integrand of each $I^{(n,m)}_{\ssc +}$ term the two cuts serve to render the $k_0$ and $l_0$ integrations trivial, i.e., they fix both the $k_0$ and $l_0$ components. Furthermore, the two individual cuts are correctly aligned (relative to their shaded sides) to form a single Cutkosky cut separating the original graph into two pieces.   
%sewn together by the remaining phase space integral. 
We, therefore, need not work out these terms any further. 

The $I^{(n)}_{\ssc +}$ terms, on the other hand, contain one cut, which amounts to fixing the $k_0$ component only. To convert them also to 2-cut terms we need to carry out the $l_0$ integration.
Consider then the general I$^{(n)}_{\ssc + }$ term (\ref{C1-cut1}).
After conveniently rerouting the $l$ momentum %performing a shift $k\to k-l$ of integration variables 
so that $q_j = k+ P_j$, $j\in Q_{II}$, we close the $l_0$ contour in the l.h.p.; the only enclosed poles are then the l.h.p. poles in propagators on internal lines in $Q_I$. 
The $l_0$ integration is again most efficiently done by replacing $\Delta(Q_I)$  in (\ref{C1-cut1}) by $\Delta_A(Q_I)$. Since the latter have no poles in the l.h.p. the result is zero, and using the relation (\ref{Gproprel1}) 
%yields
%\bml
%I^{(n)}_{\ssc +}( \{p\}_{\ssc N}) = \sum_{m\in Q_I} 
%  \int  {d^dl \over (2\pi)^d}{d^dk \over (2\pi)}\, \delta^{(d-1)}\; ({\bf k-l}) \;  \wdelta^-(q_n) \; \wdelta^+(q_m)\; {i\over k_0-l_0 + i\epsilon}  
%     \\
%\cdot  \prod_{s=1}^{N} iV_s^{(n_s)}(q_s, q_{s-1}, \{p\}_i)  
% \, \prod_{j\in Q_{II}\atop j > n}  \Delta(q_j)  \prod_{j\in Q_{II}\atop j < n}(-\Delta^*(q_j) )   
% \prod_{i\in Q_{I} \atop i\not= m}  \Delta(q_i, \tlepsilon(q_i,q_m) )    
%  \, .   \label{C1-cut2}
%\end{multline} 
results in the replacement 
\[   \prod_{i\in Q_{I}}  \Delta(q_i )  \to \sum_{m\in Q_I}  \,  \wdelta^+(q_m)\; \prod_{i\in Q_{I} \atop i\not= m}  \Delta(q_i, \tlepsilon(q_i,q_m) )  \]
in (\ref{C1-cut1}). We then proceed as above to reduce all $ \Delta(q_i, \tlepsilon(q_i,q_m) )$ propagators in $Q_I$.  
Consider first the internal lines in $Q_I$ on the shaded side of the $\wdelta^+(q_m)$ cut and use relation (\ref{proprel1a}). Here $i> m$, so $q_{i0} - q_{m0}= \sum_{k=m}^i p_{k0} > 0$ and, hence, there is no $\delta$-function contribution in (\ref{proprel1a}).  
On the unshaded side of the $\wdelta^+(q_m)$ cut use relation (\ref{proprel1}). Now $i < m$ and so $q_{i0} - q_{m0}= - \sum_{k=i}^m p_{k0} < 0$. Hence, no  $\delta$-function contribution survives in (\ref{proprel1}) either. Thus, 
after  a shift %$k\to k+l$ 
to the original momenta routing,  we can express $I^{(n)}_+( \{p\}_{\ssc N}) $ also as a sum of 2-cut 
integrals: 
\bml
I^{(n)}_{\ssc +}( \{p\}_{\ssc N})    
= \sum_{m\in Q_I} 
  \int  {d^dl \over (2\pi)^d}{d^dk \over (2\pi)}\, \delta^{(d-1)}\; ({\bf k}) \;  \wdelta^-(q_n) \; \wdelta^+(q_m)\; {i\over k_0 + i\epsilon}  
     \\
\cdot  \prod_{s=1}^{N} iV_s^{(n_s)}(q_s, q_{s-1}, \{p\}_i)  
 \, \prod_{j\in Q_{II}\atop j > n}  \Delta(q_j)  \prod_{j\in Q_{II}\atop j < n}(-\Delta^*(q_j) )   
 \prod_{i\in Q_{I} \atop i < m}  \Delta(q_i )   \prod_{i\in Q_{I} \atop i > m} (- \Delta^*(q_i ) )  
  \, .   \label{C1-cut3}
\end{multline} 
The two cuts in (\ref{C1-cut3}) are properly aligned so that they combine to form a single Cutkowsky cut separating the original graph into two pieces. 
%This, as already noted, is also the case for (\ref{C2-cut1}). 
Now, collecting the minus signs in front of the $\Delta^*$ factors on the r.h.s. of  (\ref{C1-cut3}) results in a factor $(-1)^{(n-m-1)}= -(-1)^{(n-m)}$. Since $(n-m)$ is the number of vertices on the shaded side bounded by the two cuts, we see that this factor amounts to the substitution $iV_s \to -iV_s$ with one overall minus sign left. The same holds for the corresponding minus signs on the r.h.s. of (\ref{C2-cut1}). 

We thus see that, apart from an overall minus sign, 
$I^{(n,m)}_{\ssc +}( \{p\}_{\ssc N})$, eq. (\ref{C2-cut1}), is the expression for the graph sliced by a cut within the set $Q_{II}$ of internal lines; whereas each term in $I^{(n)}_{\ssc + }( \{p\}_{\ssc N})$, eq. (\ref{C1-cut3}),  is the expression for the graph with a cut slicing the set $Q_I\cup Q_{II}$. In both cases   propagators and vertices on the unshaded side of a cut are given by the Feynman rules for $S$, whereas on the shaded side by those for $S^\dagger$.   

Substituting  (\ref{C2-cut1}) and (\ref{C1-cut3}) in (\ref{Campl3}) the result may be expressed by the graphical equation: 
 
\parbox{14cm}
{\epsfysize=2.5cm \epsfxsize=13cm \epsfbox{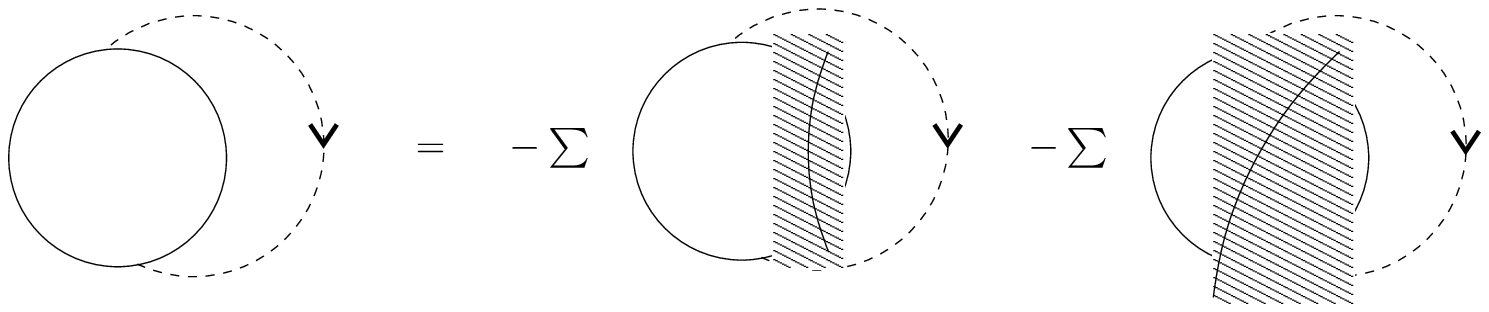}}     
  \parbox{0.7cm}{\begin{equation}\label{BCC+} 
\end{equation}}\\[0.7cm]
Again,  external legs are not depicted explicitly  in (\ref{BCC+}) (and in similar equations below).\footnote{We follow, for the most part, the graphical conventions in \cite{tHV}. \label{FtHV}}   The sums on the r.h.s. are over all cuts placed in the direction and positions indicated.
The graphical conventions incorporate the rules arrived at above, i.e., {\it propagators and vertices on the unshaded side 
of cuts are given by the Feynman rules for $S$ (i.e., propagators $\Delta(q)$, vertices $iV$), whereas on the shaded side by those for $S^\dagger$, (i.e., propagators $\Delta^*(q)$, vertices $V^* = -iV$). Energy always flows from the unshaded to the shaded side.} 

(\ref{BCC+}) is the BCC equation for the 1-loop amplitude (\ref{ampl1}) \cite{V}. It gives the  contribution to the corresponding %$\langle{\rm out} |(...)|{\rm in}\rangle$ 
matrix elements of (\ref{BCC0d}) at 1-loop level. Specifically, the term on the l.h.s. 
of (\ref{BCC+}) gives the contribution of the unity term in $S^\dagger = 1 - iT^\dagger$ in  the first term on the l.h.s. in (\ref{BCC0d}); whereas, the first term on the r.h.s. in (\ref{BCC+}) gives the $iT^\dagger$ contribution to the first term on the l.h.s. in (\ref{BCC0d}). The second term on the r.h.s. of (\ref{BCC+}) then gives the contribution to the second term on the l.h.s. of (\ref{BCC0d}). The time-ordering constraint is incorporated by the presence of the 
propagator defined in Fig. \ref{tlF4}. 
%We have thus established that the causality condition (\ref{BCC0d}) is satisfied.  

(\ref{BCC+}) was obtained for non-derivative interactions in \cite{V} via the largest time equation. We have derived it here in the presence of derivative interactions via TLD. 
%A couple of remarks concerning (\ref{BCC+}) may be made here. 
%Note that the presence of the $\wdelta^+$, $\wdelta^-$ cuts in each term on the r.h.s. of (\ref{BCC+}) 
%fix both $l_0$ and $k_0$. 
%This means that the ordering propagator (broken line) in the terms on the r.h.s. should also  be viewed as a tree propagator; 
For future reference, it  should be noted that external legs (not explicitly shown) attached to the two selected vertices joined by the ordering propagator (broken line) in (\ref{BCC+}) can be taken off shell - indeed, nowhere in the above derivation has the on-shell condition been used for them.

The equation for the opposite time-ordering is derived in a similar manner. One now starts from 
\beq 
I_-( \{p\}_{\ssc N}) =  \int {d^dl \over (2\pi)^d}{d^dk \over (2\pi)}\,  \delta^{(d-1)}({\bf k}) \; \left({-i\over k_0 - i\epsilon}\right)  \prod_{i=1}^{N} \, iV_i^{(n_i)}(q_i, q_{i-1}, \{p\}_i) 
 \, \Delta(q_i)  
  \,    \label{Campl4}
\eeq
(Fig. \ref{tlF3}(b)) 
and performs the $k_0$ integration by closing the contour in the l.h.p. The result is quickly obtained by replacing 
in (\ref{Campl4}) all $\Delta(q_i)$ by $\Delta_A(q_i)$ if $i\in Q_{II}$, closing the contour in the l.h.p. and then substituting (\ref{Gproprel1}). It amounts to the replacement  
\[   \prod_{i\in Q_{II}}  \Delta(q_i )   \to     \sum_{n\in Q_{II}}  \;   \wdelta^+(q_n)\,  \prod_{j\in Q_{II}\atop j\not= n}  \Delta(q_j, \tlepsilon(q_j,q_n))    \]
in (\ref{Campl4}), which gives  
%\bml
%I_-( \{p\}_{\ssc N}) =  \sum_{n\in Q_{II}}   \int  {d^dl \over (2\pi)^d}{d^dk \over (2\pi)}\, \delta^{(d-1)}\; ({\bf k}) \;   \wdelta^+(q_n)\,\left( {-i\over k_0 - i\epsilon}\right)  \, \Delta(q_i)   \\
%\cdot  \prod_{s=1}^{N} iV_s^{(n_s)}(q_s, q_{s-1}, \{p\}_i) 
% \, \prod_{j\in Q_{II}\atop j\not= n}  \Delta(q_j, \tlepsilon(q_j,q_n))  
%  \prod_{i\in Q_{I}}  \Delta(q_i )  
%  \, .    \label{Campl5}
%\end{multline}       %This is 
the analog to (\ref{Campl2}). We then proceed in the exact analogous manner to  (\ref{Cprod1}) -(\ref{C1-cut3}).  
We first successively reduce the 
$ \Delta(q_j, \tlepsilon(q_j,q_n)) $ propagators by use of relations (\ref{proprel1}) - (\ref{proprel1a}). This results into two groups of terms possessing cuts residing in $Q_{II}$, one group with one and the other group with two cuts. 
In the 1-cut terms the $l_0$ integration may next be performed by now closing the contour in the u.h.p., and use of (\ref{Gproprel2}). This results into a second cut $\wdelta^-(q_m)$ residing in $Q_I$, and the resulting $ \Delta(q_i, -\tlepsilon(q_i,q_m))$ propagators in $Q_I$ are next reduced by use of (\ref{proprel1b}) - (\ref{proprel1c}). The 1-cut terms are thus converted into 2-cut terms with one cut residing in $Q_{II}$ and one in $Q_I$. Again, the various factors in these reductions work out so that the Feynman rules on the shaded (unshaded) sides of the cuts are those of $S^\dagger$ ($S$). The result is\\[0.1cm]  

\parbox{14cm}
{\epsfysize=2.3cm \epsfxsize=13cm \epsfbox{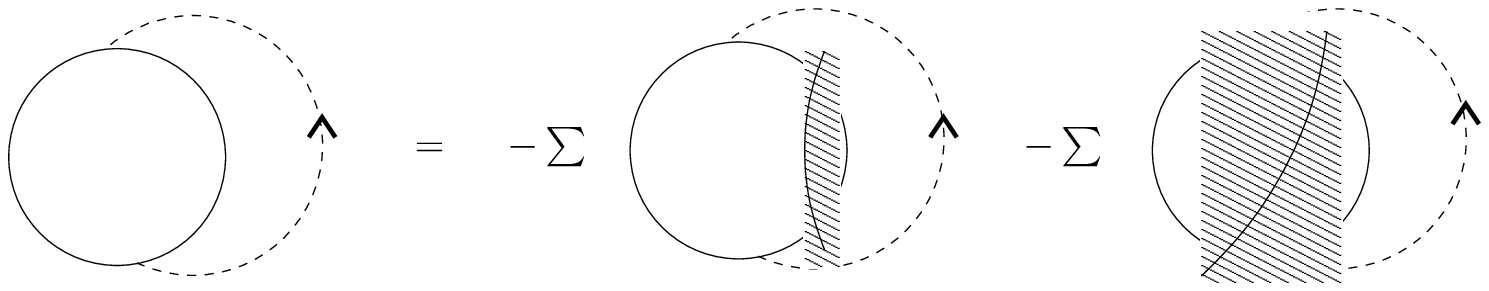}}     
%\hfill 
\parbox{0.7cm}{\begin{equation}  \label{BCC-} 
\end{equation}}\\[0.7cm]
to be contrasted with (\ref{BCC+}). It represents (\ref{BCC0d}) with the labels $x$ and $y$ interchanged.

Now, adding both time orderings (\ref{Campl1}) and (\ref{Campl4}) gives back the original amplitude. Indeed, noting that 
\beq
{1\over 2\pi} \left( {i\over k_0 + i\epsilon} -  {i\over k_0 - i\epsilon}\right)   
%=  \left({1\over \pi} {\epsilon \over k_0^2 + \epsilon^2}\right) 
= \delta(k_0)    \, ,    \label{deltak0}
\eeq 
one has 
\bea 
 I_+( \{p\}_{\ssc N})  + I_-( \{p\}_{\ssc N}) 
& = & \int {d^dl \over (2\pi)^d} \prod_{i=1}^{N} \, iV_i^{(n_i)}(q_i, q_{i-1}, \{p\}_i)  \, \Delta(q_i) \nonumber \\
 & = & A( \{p\}_{\ssc N}) \,. \label{Ctotalampl} 
 \eea
Adding the BBC equations (\ref{BCC+}) and ({\ref{BCC-}) then yields\\[0.1cm] 

\parbox{14cm}
{\epsfysize=2.3cm \epsfxsize=13.5cm \epsfbox{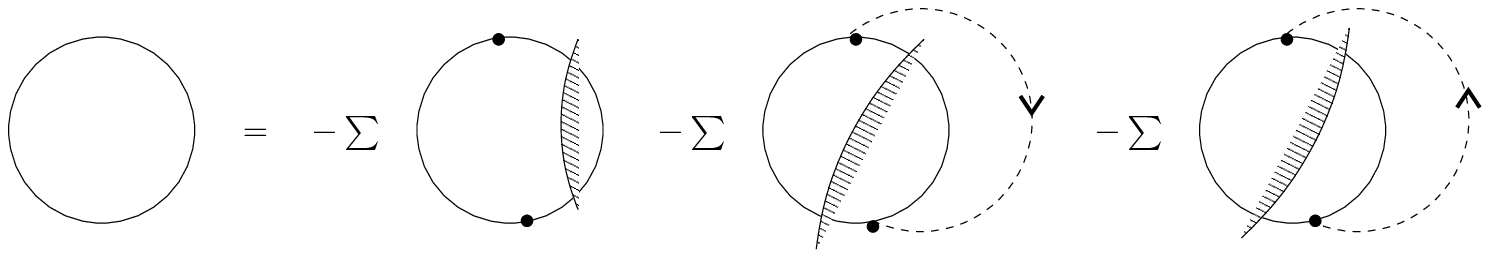}}     
\parbox{0.7cm}{\begin{equation}\label{BCCt} 
\end{equation}}\\[0.7cm]
(\ref{BCCt}) is remarkable in that it gives a representation of the complete graph in terms of cut diagrams. The sums on the r.h.s. are over all cuts placed as indicated relative to two interaction vertices (indicated by the dots) that have been picked out.  
%This representation of the complete amplitude (for non-derivative interactions) was first obtained in \cite{tHV} via the %largest time equation. 

\section{Unitarity and Microcausality\label{un} }
Once the BCC equation (\ref{BCC+}) is shown to hold in a particular theory the unitarity of the S-matrix, as well as the usual microcausality requirement follow as consequences. 
\subsection{Unitarity}  
(\ref{BCCt}) gives  an exact  representation of the amplitude in terms of Cutkosky cuts. To extract the absorptive part 
we need %to first consider the amplitude and the related 
the amplitude (\ref{Campl1}) in $S^\dagger$ rather than in $S$: 
\beq 
\hat{I}_+( \{p\}_{\ssc N}) =   \int {d^dl \over (2\pi)^d}{d^dk \over (2\pi)}\,  \delta^{(d-1)}({\bf k}) \; {i\over k_0 + i\epsilon} \, \prod_{i=1}^{N} \,(- i)V_i^{(n_i)}(q_i, q_{i-1}, \{p\}_i) 
 \, \Delta^*(q_i)  
  \, .   \label{Uampl1}
\eeq
It is now more expedient to first perform the $l_0$ integration. After a shift $k\to k-l$ rerouting the $l$ momentum we close the contour in the l.h.p., where the only poles are from the propagators in $Q_I$, and obtain: 
\bml
\hat{I}_+( \{p\}_{\ssc N}) =   \sum_{m\in Q_I}   \int  {d^dl \over (2\pi)^d}{d^dk \over (2\pi)}\, \delta^{(d-1)}\; ({\bf k-l}) \;   \wdelta^-(q_m)\, {i\over k_0-l_0 + i\epsilon}  \\
\cdot  \prod_{s=1}^{N}(- i)V_s^{(n_s)}(q_s, q_{s-1}, \{p\}_i) 
 \, \prod_{i \in Q_I \atop i \not= m} (- \Delta(q_i, \tlepsilon(q_i,q_m)) ) 
  \prod_{j\in Q_{II}}  \Delta^*(q_J )  
  \, .   \label{Uampl2}
\end{multline}
We then follow the same procedure as in the previous cases above by first successively reducing the 
$( -\Delta(q_i, \tlepsilon(q_i,q_m)))$ propagators in (\ref{Uampl2}) by use of relations (\ref{proprel1}) - (\ref{proprel1a}). This results again into two groups of terms with cuts now in $Q_I$, 
one group with one and the other group with two such cuts.   
Performing next the $k_0$ integration in the 1-cut terms  by now closing the contour in the u.h.p.  gives another cut $\wdelta^+(q_n)$ residing in $Q_{II}$. 
%and use of (\ref{Gproprel2}), 
The resulting associated  $(- \Delta(q_j, -\tlepsilon(q_j,q_n)))$ propagators in $Q_{II}$ are then reduced by use of (\ref{proprel1b}) - (\ref{proprel1c}). The  1-cut terms are thus converted into 2-cut terms with one cut residing in $Q_{II}$ and one in $Q_I$. Feynman rules on the shaded (unshaded)  sides of the cuts are those of $S^\dagger$ ($S$). The final result is\\[0.1cm]  

\parbox{14cm}
{\epsfysize=2.3cm \epsfxsize=13cm \epsfbox{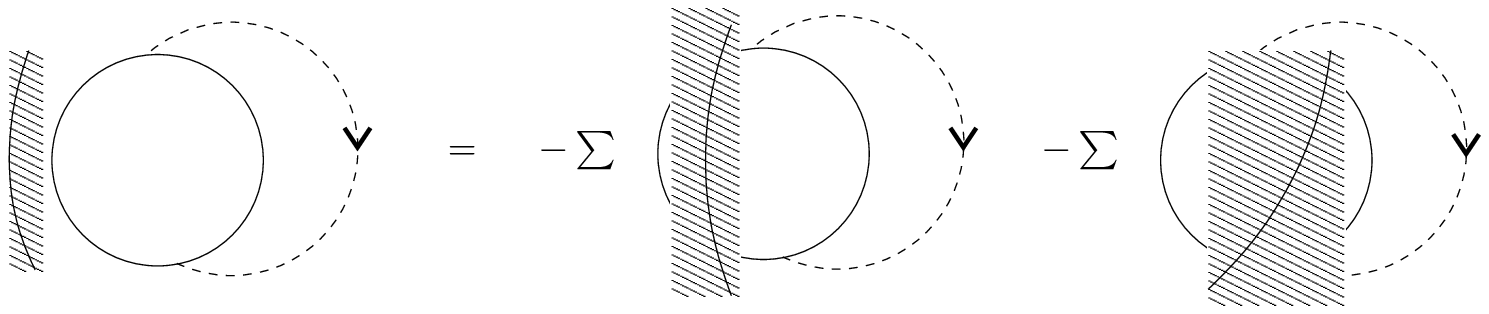}}     
%\hfill 
\parbox{0.7cm}{\begin{equation}\label{BCC+dagger} 
\end{equation}}\\[0.7cm]
(\ref{BCC+dagger}) is indeed the BCC equation (\ref{BCC0e}).

The opposite ordering 
\beq 
\hat{I}_-( \{p\}_{\ssc N}) =   \int {d^dl \over (2\pi)^d}{d^dk \over (2\pi)}\,  \delta^{(d-1)}({\bf k}) \;\left( {-i\over k_0 - i\epsilon}\right) \, \prod_{i=1}^{N} \,(- i)V_i^{(n_i)}(q_i, q_{i-1}, \{p\}_i) 
 \, \Delta^*(q_i)  
  \, .   \label{Uampl3}
\eeq
is worked out in the same manner. One now gets: 

\parbox{14cm}
{\epsfysize=2.3cm \epsfxsize=13cm \epsfbox{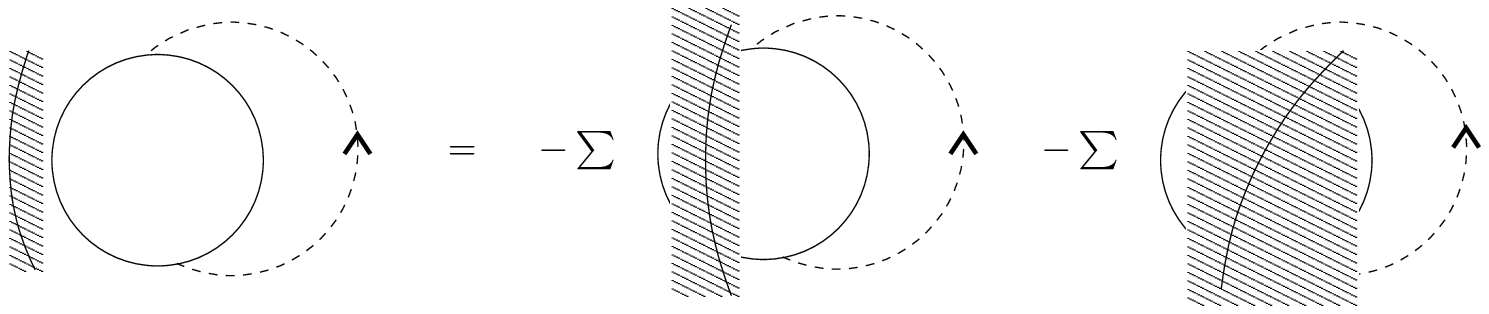}}     
%\hfill 
\parbox{0.7cm}{\begin{equation}\label{BCC-dagger} 
\end{equation}}\\[0.7cm]
 Adding now (\ref{BCC+dagger}) and (\ref{BCC-dagger}) gives:\\[0.1cm]    %the complete $S^\dagger$ amplitude 

\parbox{14cm}
{\epsfysize=2.3cm \epsfxsize=13.5cm \epsfbox{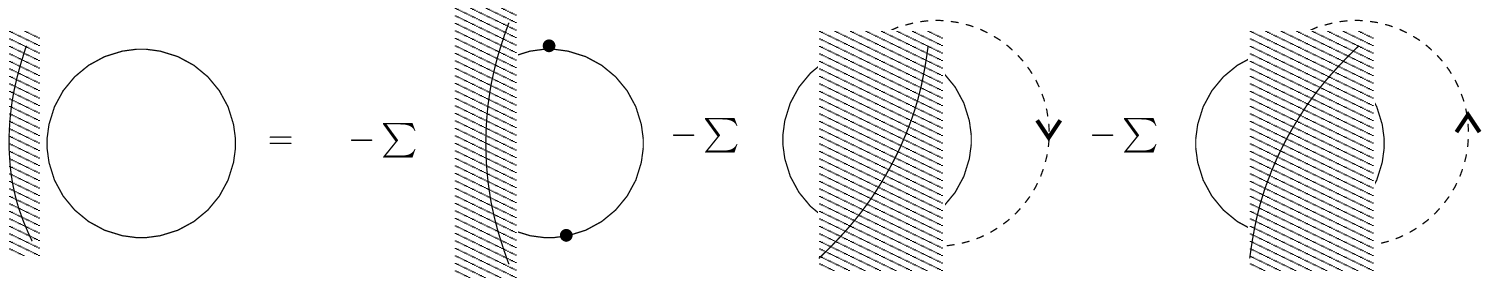}}     
%\hfill 
\parbox{0.7cm}{\begin{equation}\label{BCCtdagger} 
\end{equation}}\\[0.7cm]
This is then the BCC representation for the $S^\dagger$ amplitude. % that is the counterpart to (\ref{BCCt}). 

Adding (\ref{BCCt}) and (\ref{BCCtdagger})  gives the absorptive part:\\[0.1cm]

\parbox{14cm}
{\epsfysize=6.5cm \epsfxsize=11.5cm \epsfbox{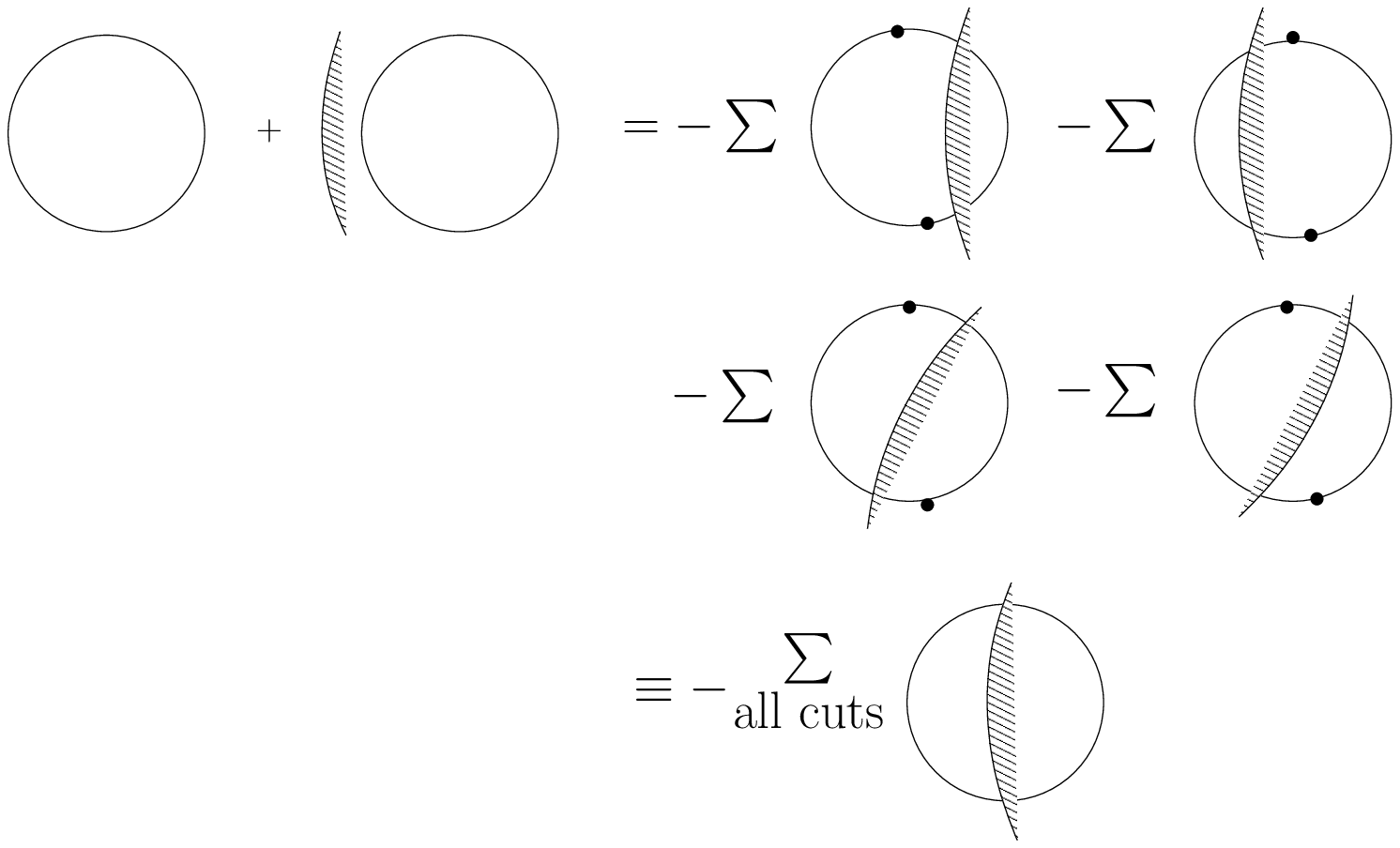}}     
\parbox[t]{0.7cm}{\begin{eqnarray} & &  \label{U1} \\
& & \nonumber \\
& &\nonumber  \\
& & \label{U2}
\end{eqnarray}}\\[0.7cm]
This is the standard unitarity equation expressing the absorptive part as a sum over Cutkosky 
cuts.\footnote{Note that graphs here represent $S$-matrix elements ({\it not} $T$-matrix elements); hence the plus sign on the l.h.s. (cf. footnote \ref{FtHV}).\label{Ueq.}} Indeed, on the r.h.s. of (\ref{U1}) all cut positions relative to the original two chosen vertices are now present, so these two vertices are no longer distinguished from all the other vertices and, hence, no longer relevant. Note in this connection 
that reversing the cut orientation in the first two terms on the r.h.s. would give a vanishing contribution 
by energy conservation; hence, both cut orientation in the first two terms could be included as well. As a result the r.h.s. can be given simply as in (\ref{U2}), i.e., the sum over all cuts, including cuts that are not allowed kinematically and so will not contribute. The general rule is easily seen to be (cf. \cite{tHV}) that allowed non-vanishing cuts are such that the shaded (unshaded) side includes at least one vertex having an outgoing (ingoing) leg attached. 

Though obvious, it is perhaps still worth pointing out that (\ref{U2}), i.e.,  the statement $SS^\dagger=1$,  holds, with the above assumptions concerning Feynman rules (in particular, hermiticity of interaction Lagrangians), even in the  presence of some propagators possessing negative residues. Such propagators may be the result of regularization, as in the case of PV regulators, or the unphysical excitations in gauge theories. 
In such a case to ensure {\it  physical} unitarity, i.e., that the statement (\ref{U2}) holds separately also for the restriction of $S$ to the sector comprising only positive residue propagators, requires additional considerations, e.g., taking an appropriate limit, or taking into account constraints such as the Ward identities.

\subsection{Microcausality}

The general BCC equation (\ref{BCC+}) implies, in particular, that  the usual microcausality condition is satisfied.  It suffices to consider the case $N=2$. 
%, i.e., the 2-point function which in fact corresponds to the 
%commonly stated requirement of commutativity of two operators (fields) at spacelike separation. 
For $N=2$ there are only two vertices present with incoming and outgoing momentum $p$. To extract the microcausality condition
subtract the sum of (\ref{BCC-} and (\ref{BCC+dagger}) from the sum of (\ref{BCC+}) and (\ref{BCC-dagger}). 
The result is (cf. (\ref{BCC0mc1})):
\nopagebreak
\parbox{14cm}
{\epsfysize=3.5cm \epsfxsize=12.5cm \epsfbox{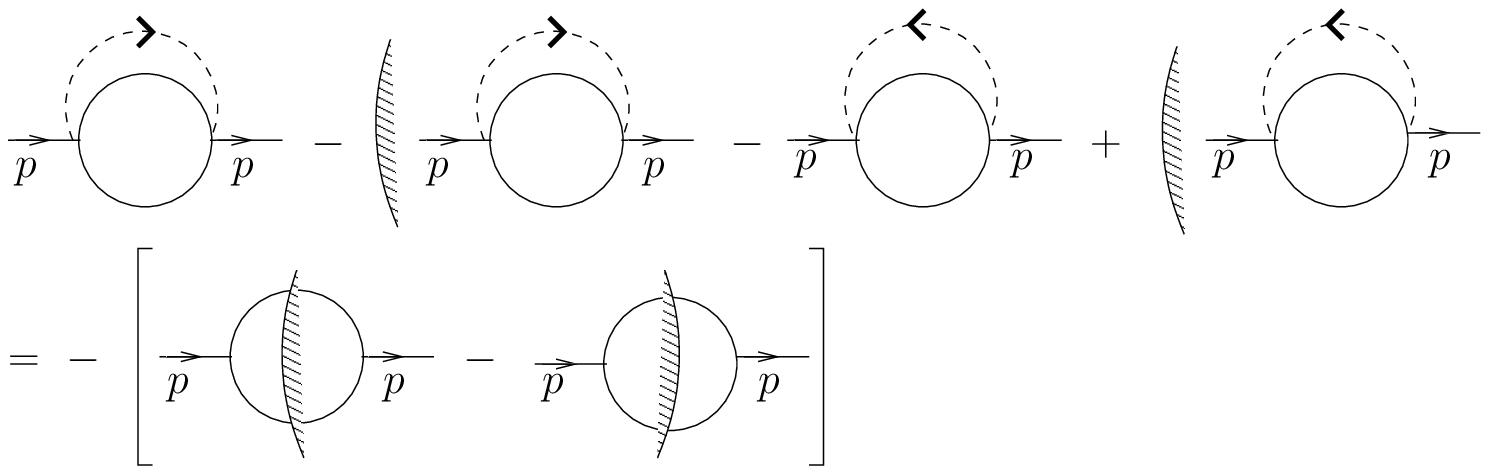}}     
\parbox[t]{0.7cm}{\beq  \label{MC1} 
\eeq}\\[0.7cm]
As noted above (cf. remarks following (\ref{BCC+})), the legs attached to the two vertices may be taken off shell. We then revert 
to coordinate space by multiplying each term by $\exp ip(x-y)$ and integrating over $p$.  
The r.h.s. represents the 1-loop contribution to the commutator of vertex operators inserted at $x$ and $y$. It is now straightforward to verify directly that it indeed vanishes for spacelike $(x-y)$. One has
\begin{align}
& \parbox{14cm}
{\epsfysize=1.7cm \epsfxsize=4.5cm \epsfbox{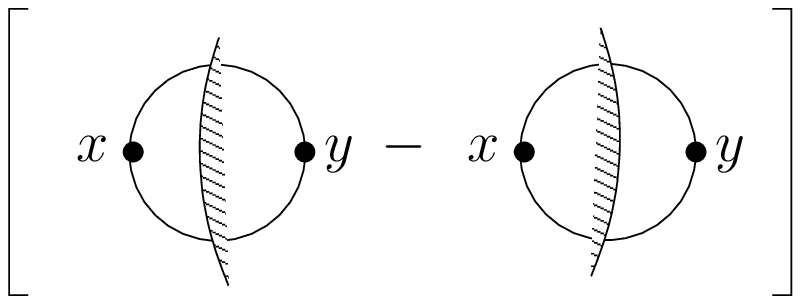}}  \nonumber  \\ 
&= \int {d^d p\over (2\pi)^d} {d^d l\over (2\pi)^d} \, V_1(l+p,l, p) V_2(l, l+p,-p)  \left[ \wdelta^-(l) \wdelta^+(l+p) - \wdelta^+(l) \wdelta^-(l+p) \right]
e^{ip(x-y)}  \label{MC2} \\ 
 &=\int {d^d p\over (2\pi)^d} {d^d l\over (2\pi)^d} \, V_1(l+p,l, p) V_2(l,l+p,-p)\,   \wdelta^-(l) \wdelta^+(l+p) e^{-i{\bf p \cdot (x-y)}} 2i \sin [p_0 (x^0- y^0)]
\, .\label{MC3} 
\end{align}
This clearly vanishes for $(x^0-y^0) =0$ and, hence, by relativistic invariance (manifest in (\ref{MC2})) for any spacelike $(x-y)$. The l.h.s of (\ref{MC1}) vanishes then by the equality, but this can, of course, be similarly verified directly. 

By the same manipulation one may write down the analog of (\ref{MC1}) for general $N$, the r.h.s. now being the expression for commuting two of the operators among a string  of N vertex operators. It represents the corresponding matrix element of (\ref{BCC0mc1}). 
Its vanishing for spacelike separations is the general statement of microcausality (cf. \cite{SW}).

\section{Multi loops \label{ml}}  

The extension to multiloop graphs is illustrated by the 2-loop graph in Fig. \ref{2lF5}(a). The associated graph with ordering between two vertices is shown in Fig. \ref{2lF5}(b) and given by the amplitude 
\beq
J_+(\{p\}) = \int {d^d l_1\over 2\pi^d} {d^d l_2\over 2\pi^d} {d^d k\over 2\pi}\,   \delta^{(d-1)}({\bf k }) \, {i\over k_0 + i\epsilon} 
\prod_{s=1}^{N+1} iV_s^{(n_s)} (\{q\}_s, \{p\}_s) \prod_{i\in Q_I\cup Q_{II}\cup Q_{III} } \Delta(q_i) \, .   \label{2lCampl1}
\eeq
\begin{figure}[ht]
\begin{center}
\includegraphics[width=0.6\textwidth]{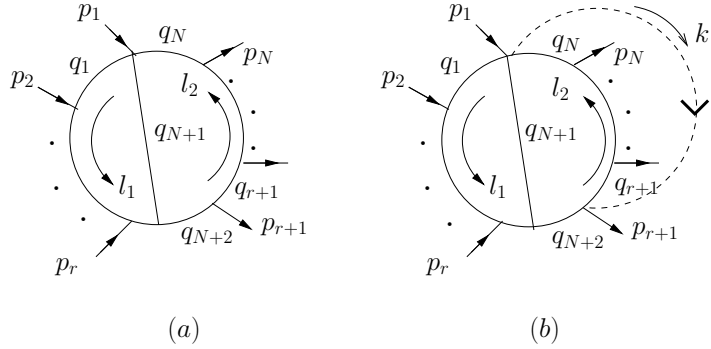}
\end{center}
\caption{(a) Two loop graph; (b) same graph with causal ordering of two vertices.  \label{2lF5}}
\end{figure}
Here we defined 
%$Q_I = \{ q_1, q_2,\cdots , q_r\}$, $Q_{II} = \{ q_{r+1}, q_{r+2},\cdots , q_N\}$, and $Q_{III} = \{ q_{N+1}, q_{N+2}\}$.  
\[ Q_I = \{ q_1, q_2,\cdots , q_r\} \, , \quad Q_{II} = \{ q_{r+1}, q_{r+2},\cdots , q_N\} \, , \quad Q_{III} = \{ q_{N+1}, q_{N+2}\} \, . \]
The internal line momenta are then 
\bea
q_i & = & l_1 + P_i \, , \qquad  P_i = \sum_{k=1}^i p_k , \quad i\in Q_I   \nonumber \\
 q_i & = & l_1 + l_2 + k + P_i \, , \qquad  P_i =P_r -  \sum_{k=r+1}^i p_k ,  \quad i\in Q_{II}   \nonumber \\
 q_{N+1} &=&  l_2 \;, \qquad q_{N+2} = l_1+ l_2 +P_r     \, . \label{2lintmom1} 
\eea
We apply the procedure used in the 1-loop cases above. We first do the $k_0$ integration closing the contour in the u.h.p. This results in the replacement 
\beq 
\Delta(Q_{II}) = \prod_{j\in Q_{II}} \Delta(q_j) \quad  \to \quad \wDelta^-(Q_{II}) 
= \sum_{n \in Q_{II}} \wdelta^-(q_n) \prod_{j\in Q_{II} \atop j\not= n} \Delta\big(q_j, -\tlepsilon(q_j,q_n)\big)    \label{2lCampl2}
\eeq
in (\ref{2lCampl1}). The propagators $\Delta(q_j, -\tlepsilon(q_j,q_n))$ in (\ref{2lCampl2}) are next reduced by use of the relations (\ref{proprel1b}) and (\ref{proprel1c}). The result is the exact analog to  (\ref{Campl3}): 
\begin{subequations} \label{2lCampl3}
\bal
J_+( \{p\}_{\ssc N}) 
%& = & I_{\rm 1-cut}( \{p\}_{\ssc N}) + I_{\rm 2-cut} ( \{p\}_{\ssc N})  \nonumber \\
&=  \sum_{n\in Q_{II}} J^{(n)}_{\ssc +}( \{p\}_{\ssc N})   + \sum_{n,m \in Q_{II}\atop n>m} 
J^{(n,m)}_{\ssc +}( \{p\}_{\ssc N})   \, , \label{2lCampl3a}\\
=& \parbox{12cm}
{\epsfysize=2.2cm \epsfxsize=9.5cm \epsfbox{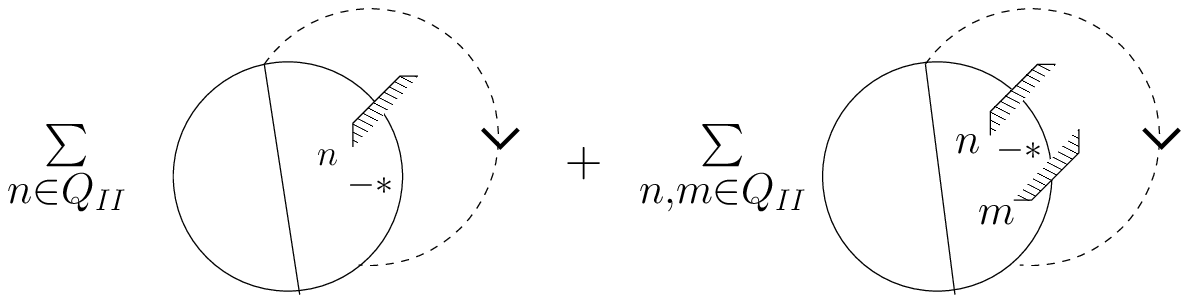}}     \label{2lCampl3b}
\end{align}
\end{subequations}
%where now $J^{(n)}_{\ssc +}( \{p\}_{\ssc N})$ and $J^{(n,m)}_{\ssc + }( \{p\}_{\ssc N}) $ are graphically represented in %Fig. . 
The $J^{(n,m)}$ terms already contain two cuts which can be united into a single cut cutting through the graph. They then need not be worked out any further.

The $J^{(n)}_{\ssc +}$ terms possess one cut. We now perform the $l_{20}$ integration in each $J^{(n)}_{\ssc +}$ term closing the contour in the l.h.p. 
After a shift $k \to k - l_1-l_2$, conveniently  rerouting loop momenta through the vertex ordering propagator, the result of this integration, by another application of the tree-loop relation (\ref{tld}),  is the replacement 
\bml
\Delta(Q_{III}) = \prod_{j\in Q_{III}} \Delta(q_j)       
\quad \to \quad  \wDelta^+(Q_{III}) = 
\wdelta^+(q_{N+1}) \Delta\big(q_{N+2}, \tlepsilon(q_{N+2},q_{N+1})\big)  \\ 
 + \wdelta^+(q_{N+2}) \Delta\big(q_{N+1}, \tlepsilon(q_{N+1},q_{N+2})\big) 
% = \sum_{n \in Q_{III}} \wdelta^-(q_n) \prod_{j\in Q_{III} \atop j\not= n} \Delta(q_j, \tlepsilon(q_j,q_n))    
\, .  \label{2lCampl4} 
\end{multline}  
Now $(q_{N+2} - q_{N+1}) =  l_1 + P_r$. Hence, the $\tlepsilon\,$'s in (\ref{2lCampl4}) depend on the loop momentum $l_1$, which makes the pole locations depend on an integration variable. This is a new feature that enters at the 2-loop level and beyond. It renders further reduction of propagators in (\ref{2lCampl4}) and  subsequent integration over $l_1$ problematic. To disentangle this dependence  we use relation (\ref{Gproprel3}) to transform the r.h.s. in (\ref{2lCampl4}). Since $\wDelta^+(Q_{III})= \wDelta^+(q_{N+1} \cup q_{N+2})$, (\ref{Gproprel3}) gives 
\beq
\wDelta^+(Q_{III}) =  - \wdelta^+(q_{N+1}) \wdelta^+(q_{N+2}) + \wdelta^+(q_{N+1})\Delta(q_{N+2}) + 
\wdelta^+(q_{N+2})\Delta(q_{N+1})   \, . \label{2lRrel1} 
\eeq
Inserting  (\ref{2lRrel1}) the result for $J^{(n)}_{\ssc +}$ is given graphically by \\[0.1cm]

\parbox{14cm}
{\epsfysize=2.5cm \epsfxsize=13.5cm \epsfbox{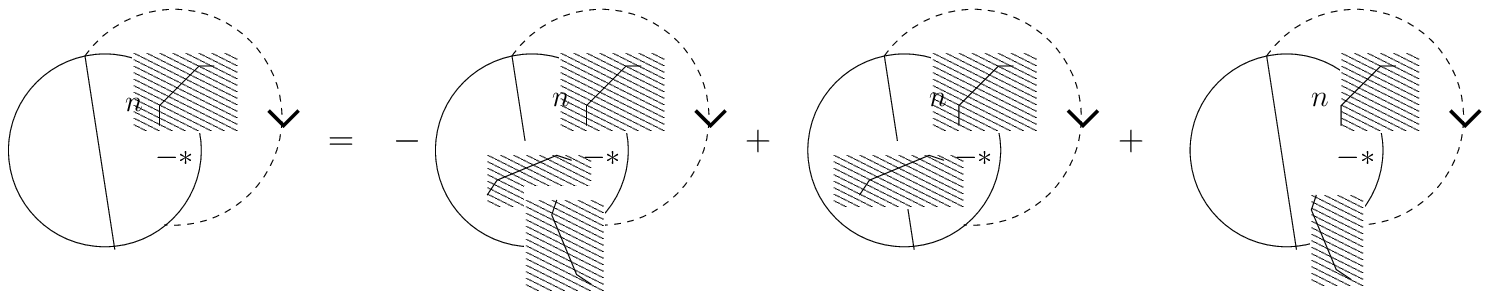}}     
\parbox[t]{0.7cm}{\beq  \label{2lCampl5} 
\eeq}\\[0.7cm]        
where the three terms correspond  to the three terms in (\ref{2lRrel1}). 
%The third term on the r.h.s. of \ref{2lCampl5has two cuts that may be united into a single cut.  
In the second term on the r.h.s. of (\ref{2lCampl5}) all propagators depending on $l_1$ are now $\Delta$ propagators so that we may perform the $l_{10}$ integration. Closing the contour in the l.h.p. this amounts again to an application of (\ref{tld}) and results in the replacement  
\bml 
\Delta(Q_I) \Delta(q_{N+2}) \quad  \to \quad \wDelta^+(Q_I\cup q_{N+2}) = 
 \sum_{m \in Q_I} \wdelta^+(q_m) \, \Delta\big(q_{N+2}, \tlepsilon(q_{N+2},q_m)\big)  \prod_{i\in Q_I \atop i\not= m}  \Delta\big(q_i, \tlepsilon(q_i,q_m)\big) 
   \\
  +   \wdelta^+(q_{N+2})  \prod_{i\in Q_I} \Delta\big(q_i, \tlepsilon(q_i,q_{N+2})\big)    \label{2lCampl6}
  \end{multline} 
The propagators on the r.h.s. in (\ref{2lCampl6}) are next reduced as before by use of the relations (\ref{proprel1}) and (\ref{proprel1a}). We note that, for the terms with the $\wdelta^+(q_m)$ cut in (\ref{2lCampl6}), one has $i,m \in Q_I$ and $(q_{i0} - q_{m0}) = (P_{i0} - P_{m0}) > 0$ $ (< 0)  $ if $i> m$ ($i< m$), i.e.,  on the shaded (unshaded) side of the  cut.  Also, 
\[ q_{(N+2)0} - q_{m0} = l_{20} +  (P_{r0} - P_{m0})  = \omega_{{\bf l}_2} +  (P_{r0} - P_{m0}) > 0 \]
since $q_{(N+1)0} = l_{20} = \omega_{{\bf l}_2}$ due to the pre-existing cut on the $q_{N+1}$ line in the second  term   
in (\ref{2lCampl5}). For the term with the $\wdelta^+(q_{N+2})$ cut in (\ref{2lCampl6}) one has $(q_{i0} - q_{(N+2)0})= - 
\omega_{{\bf l}_2} -  (P_{r0} - P_{i0})  < 0$. As a result, use of (\ref{proprel1}) and (\ref{proprel1a}) gives \\[0.1cm]

\parbox{14cm}
{\epsfysize=2.5cm \epsfxsize=12.5cm \epsfbox{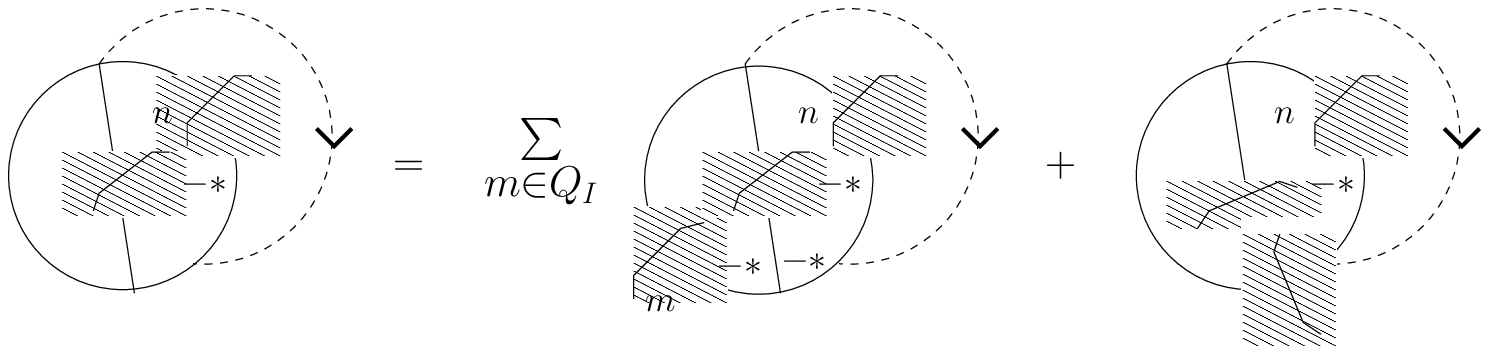}}     
\parbox[t]{0.7cm}{\beq  \label{2lCampl7} 
\eeq}\\[0.7cm]   
Substituting (\ref{2lCampl7}) then into (\ref{2lCampl5}) one gets \\[0.1cm]

\parbox{14cm}
{\epsfysize=2.5cm \epsfxsize=12.5cm \epsfbox{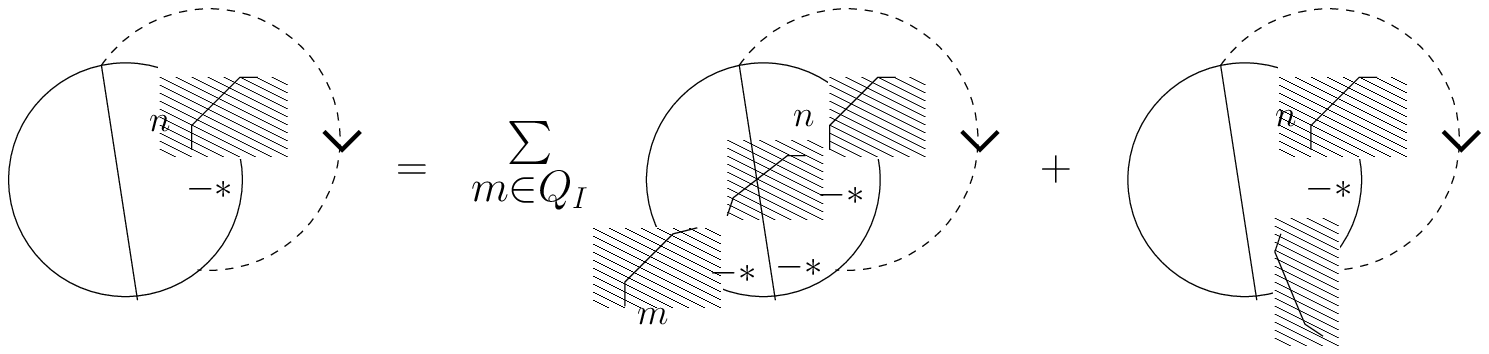}}     
\parbox[t]{0.7cm}{\beq  \label{2lCampl8} 
\eeq}\\[0.7cm]   
The cuts on the r.h.s can now be united into single cuts slicing through the graphs. 
Again, as easily verified, the minus signs in front of $-\Delta^*$ propagators on the shaded sides of cuts amount to the replacement $iV_s \to -iV_s=V_s^*$ of the vertices in each graph with one overall minus sign left over. 
Inserting (\ref{2lCampl8}) into (\ref{2lCampl3}) then we obtain the BCC for the 2-loop amplitude in Fig. \ref{2lF5}(a): 
\bml
\parbox{14cm}
{\epsfysize=2.1cm \epsfxsize=15cm \epsfbox{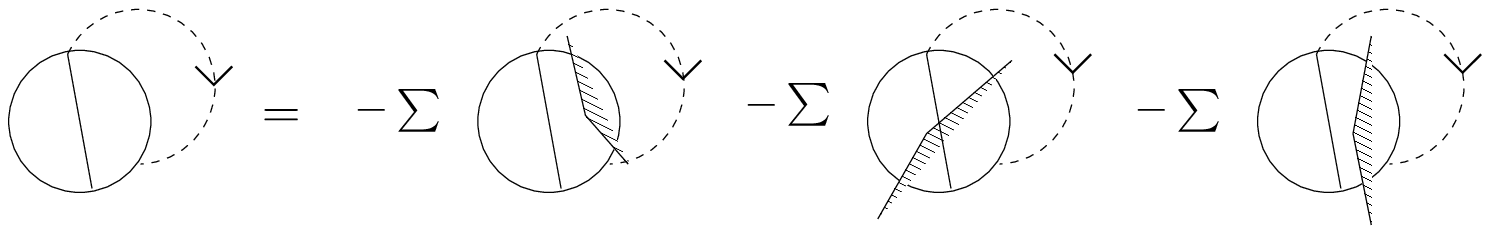}}   \\
  \label{2lBCC+} 
%\parbox[t]{0.7cm}{\beq  \label{BCC+} 
%\eeq}\\[0.7cm] 
%\beq 
%BCC +   \label{2lBCC+}
%\eeq
\end{multline}
At this point, following \cite{tHV}, we may as well represent (\ref{2lBCC+})  by (\ref{BCC+}) viewing the circle in the latter equation as a ``blob" standing for the 2-loop amplitude. Indeed, with this graphical convention (\ref{BCC+}) unambiguously specifies all the cuts in (\ref{2lBCC+}). Combining with the BCC equation with opposite vertex ordering all the consequences derived in the 1-loop case are now obtained for the 2-loop case in exactly analogous manner. In particular, one has the representation (\ref{BCCt}) for the complete 2-loop graph. 
 
In a similar manner, and, in particular, with the help of identities such as  (\ref{Gproprel3}) and its straightforward generalizations to more than two sets of internal lines, one can treat higher loops. There is one further complication  though, that may occur in higher loops. This is the possibility of two or more identical propagators, which leads to poles of order higher than first. The way to deal with this by partial integration converting back to first order poles has been
presented in \cite{BBDMR}.

\section{Conclusion}
Starting from the TLD representation of a general loop graph we have obtained the BCC equation (\ref{BCC+}) and, as a further consequence, the BCC representation of a graph in terms of Cutkosky cuts given by (\ref{BCCt}). Combining this representation with the corresponding equation for the graph in $S^\dagger$ allows one to extract the absorptive part and obtain the unitarity relation (\ref{U2}). The approach is very general and can be applied to any theory given in terms of a set of Feynman rules. 

These results hold, in particular, for general local derivative interaction vertices derived from a real, or, more generally, hermitian  interaction Lagrangian (provided, of course, the original graph can be properly regularized). One may also note in this connection that spin only results in the modification of the propagator numerators by momentum polynomials (entire functions) and makes no actual difference in the derivation. 
%It is indeed important to have such a generally applicable method. 
This is in contrast to the approach via the largest time equation \cite{V} which runs into special difficulties in the presence of derivative interactions and/or higher spin. Thus, for example, explicitly checking the causality constraints on a diagrammatic basis in gravitational amplitudes by this spacetime-based method quickly becomes problematic. 
The present method does not encounter any such special difficulties. This is due to the fact that the starting point provided by the TLD relation is formulated directly in momentum space. The momentum space formulation allows a direct physical interpretation of the singularity structure of  Feynman integrals \cite{CN}, which indeed underlies the results in this paper. 

Apart from the general field-theoretic significance of the above results, the BCC representation of a graph in terms of lower order cut graphs may have several particular applications. 
It can and has  been employed in the past for the recursive isolation of subdivergences in perturbative renormalization (\cite{TT1}, also, cf. some discussion in \cite{tHV}). The general approach of reconstructing graphs from cuts has, of course, played an important role in contemporary research in the computation of amplitudes in gauge theories. 
%the many advances in the computation of amplitudes in gauge theories in recent years. 
The analytic unitarity method \cite{Betal} and the more recent numerical unitarity methods (cf. \cite{Aetal} and extensive references therein) have been very effective in the computation of one-loop and some two-loop amplitudes, and  are being actively pursued in QCD (multi-leg) two- and higher loop implementations. In addition, the TLD relation itself has been directly used for evaluation of loop diagrams  \cite{ChBDR}, \cite{BChDR}. 
The BCC representation provides a ready-made representation which, furthermore, can be used iteratively, as it can be applied to the lower-order cut graphs themselves. Whether these features allow it to be used as an effective tool for the direct evaluation of amplitudes is a very interesting open question to be investigated. In any case it can provide a useful constraint in the construction of amplitudes through lower order cuts.

A further interesting question is the extent to which our considerations here can be extended to interactions of infinite derivative order. Such vertices occur in string field theory and nonlocal gravity models. To deal with such nonlocal interactions using the approach in this paper would require some generalization of the derivation of the starting TLD relation. Indeed, this derivation presently relies on one's ability to close contours at infinity in order to apply Cauchy's theorem. This cannot generally be done for nonlocal interactions except in some special cases. We hope to return to this question elsewhere.

\end{document}